\providecommand{\adsurl}[1]{\href{#1}{ADS}}

\documentclass[iop,twocolumn]{emulateapj} 
\usepackage{graphics,graphicx,epsf,natbib}
\usepackage{subfigure}
\usepackage[colorlinks=true,linkcolor=blue,anchorcolor=blue,citecolor=blue,urlcolor=blue]{hyperref}
\usepackage{color}
\usepackage{graphicx}
\usepackage{amssymb,amsmath}

\usepackage{natbib}

\slugcomment{Submitted to ApJ 08 Oct 2015; Accepted 11 Jan 2016}

\shorttitle{CLASH Cluster Morphologies}
\shortauthors{Donahue et al.}

\begin{document}


\title{The Morphologies and Alignments of Gas, Mass, and the Central Galaxies of {\it CLASH} Clusters of Galaxies}


\author{Megan Donahue\altaffilmark{1,2}, 
             Stefano Ettori\altaffilmark{3}, 
             Elena Rasia\altaffilmark{4,5},
             Jack Sayers\altaffilmark{6},
             Adi Zitrin\altaffilmark{6,10},
             Massimo Meneghetti\altaffilmark{3,6},
             G. Mark Voit\altaffilmark{1}, Sunil Golwala\altaffilmark{6}, Nicole Czakon\altaffilmark{7}, Gustavo Yepes\altaffilmark{8},
			 Alessandro Baldi\altaffilmark{1}, Anton Koekemoer\altaffilmark{9}, Marc Postman\altaffilmark{9}}
\altaffiltext{1}{Physics and Astronomy Dept., Michigan State University, East Lansing, MI, 48824 USA}
\altaffiltext{2}{donahue@pa.msu.edu }
\altaffiltext{3}{Osservatori Astronomico di Bologna, Via Ranzani 1, 40127 Bologna, Italy; INFN, Sezione di Bologna, viale Berti Pichat 6/2, I-40127 Bologna, Italy}
\altaffiltext{4}{Department of Physics, University of Michigan, Ann Arbor, MI, 48109, USA}
\altaffiltext{5}{INAF-Osservatorio Astronomico of Trieste, via Tiepolo 11, 34121 Trieste, Italy}
\altaffiltext{6}{Division of Physics, Math, and Astronomy, California Institute of Technology, Pasadena, CA 91125}
\altaffiltext{7}{Institute of Astronomy and Astrophysics, Academia Sinica, No. 1, Sec. 4, Roosevelt Rd, Taipei 10617, Taiwan, R. O. C.}
\altaffiltext{8}{Departamento de F\'isica Te\'orica, Universidad Aut\'onoma de Madrid, Cantoblanco, 28049, Madrid, Spain}
\altaffiltext{9}{Space Telescope Science Institute, 3700 San Martin Drive, Baltimore, MD, 21218, USA}
\altaffiltext{10}{Hubble Fellow}

\begin{abstract}

Morphology is often used to infer the state of relaxation of galaxy clusters. 
The regularity, symmetry, and degree to which a cluster is centrally concentrated inform quantitative measures of cluster morphology.
The Cluster Lensing and Supernova  survey with Hubble Space Telescope (CLASH) used weak and 
strong lensing to measure the distribution
of matter within a sample of 25 clusters, 20 of which were deemed to be ``relaxed" based on 
their X-ray morphology and alignment of the X-ray emission with the BCG. Towards a quantitative characterization
of this important sample of clusters, we present uniformly estimated  
X-ray morphological statistics for all 25 CLASH clusters. We compare X-ray morphologies of CLASH clusters with those
identically measured for a large sample of simulated clusters from the MUSIC-2 simulations, selected by mass. 
We confirm a threshold in X-ray surface brightness concentration of $C\gtrsim 0.4$ for cool-core clusters, where $C$ is
the ratio of X-ray emission inside 100 $h_{70}^{-1}$ kpc compared to inside 500 $h_{70}^{-1}$ kpc.
We report and compare morphologies of these clusters  
inferred from Sunyaev-Zeldovich Effect (SZE) maps of the hot gas and in from projected mass maps based on strong and weak lensing.
We find a strong agreement in alignments of the orientation of major axes for 
the lensing, X-ray, and SZE maps of nearly all of the CLASH clusters at radii of 500 kpc (approximately
1/2 $R_{500}$ for these clusters). 
We  also find a striking alignment of clusters shapes at the 500 kpc scale, as measured with X-ray, SZE, and
lensing, with that of the near-infrared stellar light at 10 kpc scales for the 20 ``relaxed" clusters. This strong alignment
indicates a powerful coupling between the cluster- and galaxy-scale galaxy formation processes.
\end{abstract}

\keywords{galaxies: clusters: intracluster medium }

\section{Background: Cluster Morphology}

Clusters of galaxies represent the largest gravitationally bound systems in the universe, 
and their gravitational potentials are dominated by dark matter ($\sim 85\%$) \citep[e.g.][]{2005RvMP...77..207V,2006ApJ...640..691V}. 
The projected mass density of a cluster can be inferred from measurements of the distortion, statistical shear and magnification that gravitational lensing induces in background galaxies 
\citep[e.g.][]{1990ApJ...349L...1T}.
The gravitational potential also binds hot, X-ray emitting intracluster gas to the cluster \citep[e.g.][]{1982ARA&A..20..547F}. 
In a relatively relaxed cluster, the shape and depth of the gravitational potential and the entropy distribution of the gas completely determine its distribution in space and temperature \citep{2001Natur.414..425V}. 
X-ray and Sunyaev-Zeldovich (\citealt{1972CoASP...4..173S})  observations of relaxed clusters therefore trace the shape, centroid, and slope of the gravitational potential, but dynamical interactions can produce shocks or pressure waves that disturb the gas and complicate the relationship between the gas distribution and the gravitational potential \citep{2008ApJ...680...17W}.
Historically, the locations and redshifts of the cluster galaxies themselves have been used to infer a projected model
for the distribution of matter in the cluster, which then can be compared to a three-dimensional model inferred from 
the observations and analysis of the hot, X-ray emitting gas \citep{1983AJ.....88..697K,1989ApJ...336...77F}. 
By combining multiple probes of the matter distribution in galaxy clusters we can 
minimize the dependence of our mass inferences on assumptions such as isotropy, symmetry, or hydrostatic equilibrium.

There is a rich history of classifying clusters of galaxies according to visual morphology.  
\citet{1958ApJS....3..211A} did not provide morphological classifications of the clusters in his famous catalog, but Zwicky and his collaborators classified clusters in terms of their central concentration as compact, medium compact or open \citep[e.g.,][]{1961cgcg.book.....Z}. 
\citet{1961PNAS...47..905M} divided a sample of 20 nearby Abell clusters into two classes, based on the types of galaxies in the cluster.
This notion later developed into the Bautz-Morgan classification system \citep{1970ApJ...162L.149B}, which distinguishes clusters by the presence of a dominant central galaxy (type I), the dominance of ellipticals but no single BCG (type II), and the rest (type III). 
Correlations between the concentration, richness of a cluster, and its Bautz-Morgan type suggested
a connection between the dynamic state of a cluster and its appearance. \citep[See review in][]{1977ARA&A..15..505B}.)

Because of the potential connection to cosmological studies, there was great interest in the 1990s 
in trying to find robust methods of constraining the total matter density of the Universe, the primordial power spectrum, or other cosmological parameters using morphological cues from
clusters of galaxies  \citep[e.g.,][]{1982A&A...107..338B,1995MNRAS.273...30D,1997ApJ...479..632S}. These efforts ultimately proved unsatisfactory, in part because the galaxy counts used to define cluster shapes are prone to systematic uncertainties, even at relatively low redshifts. 
A more extended review of the morphological properties of clusters of galaxies 
can be found in \citet{2013AstRv...8a..40R}. 

X-ray imaging of the hot gas in clusters of galaxies provides a more straightforward
means to reveal substructure and cluster shape. The intracluster gas represents some $\sim 85\%$ of the baryons in the cluster.
The X-ray emission from the intracluster gas, proportional to $n_e^2$ and a weaker
function of temperature, is not as affected by projections
and shot noise. Early X-ray images of clusters of galaxies from the Einstein satellite and their classification suggested a further connection between the 
prominence and centrality of the BCG and the dynamical state the system \citep{1982ARA&A..20..547F, 1984ApJ...276...38J, 1999ApJ...511...65J}. X-ray imaging of the 
cluster gas in single Abell clusters revealed substructures undetected in the galaxy distribution on the sky but 
visible with dense redshift sampling \citep[e.g.,][]{1983ApJ...264..356B}.  Identification of substructure in clusters may allow
for increased purity of a sample of relaxed clusters.
For example, \citet{2015MNRAS.449..199M} imposed rigorous 
selection based on cluster dynamical state in order to tightly constrain the cosmological evolution of the gas mass fraction of relaxed clusters.
In this paper, we will discuss X-ray 
measurements made with the Advanced CCD Image Spectrometer (ACIS) on board the Chandra X-ray Observatory, which is capable
of sub-arcsecond resolution of structures in X-ray emission from 10 million K gas.

The hot intergalactic gas in a cluster of galaxies can also be studied via its scattering signature on
the cosmic microwave background (CMB) \citep{1972CoASP...4..173S}. 
This Sunyaev-Zeldovich effect (SZE) measures the frequency-dependent shift in the 
CMB radiation intensity, induced by the interaction of the CMB photons with the hot intracluster electrons \citep{1965PhFl....8.2112W}.
The SZE scales with the electron pressure ($n_e T_e$) integrated along the line of sight, and therefore provides a 
gas measurement that is nearly independent of the X-ray estimates, although often X-ray spectra are used to 
constrain gas temperatures. 
While the earliest measurements of this effect came from using beam-switching techniques
with a single dish scanning across the cluster \citep[e.g.,][]{1978MNRAS.185..245B}, SZE images were enabled by the use of interferometric arrays of radio telescopes, with most of the elements packed {\ closely together  to achieve short baselines to 
tease out the extended (arcminute-scale) signal \citep[See review in ][]{2002ARA&A..40..643C}. 
Currently, most SZE images are collected using large-format bolometric cameras, which are better than X-ray images at recovering 
emission from gas at large radii to and beyond the cluster virial radius \citep[e.g., ][]{2010ApJ...716.1118P, Sayers2013}
 and are weighted more heavily towards larger radii due to the weaker 
dependence of the SZE signal on electron density.

The association of radio halos with clusters displaying irregular X-ray morphology provided support for the idea that a cluster's X-ray appearance can be used to discriminate between regular (relaxed) clusters and disturbed (dynamically active) ones (\citealt{2013AstRv...8a..40R} and references therein).
\citet{2010ApJ...721L..82C} used quantitative methods applied to the X-ray surface brightness distribution, such as the measure of the centroid shift, the concentration parameter and the third-order power ratio, to characterize substructures in a statistical sample of 32 X-ray luminous galaxy clusters, with available radio (GMRT and/or VLA) observations. 
They showed that giant radio halos prefer to be associated with dynamically disturbed galaxy clusters, characterized by high values of the X-ray centroid shift and third power ratio moment, and low values of the concentration parameter. 
\citet{2015arXiv150603209C}, by studying a mass-selected sample of 75 galaxy clusters from the Planck SZE catalogue in the redshift range $0.08 < z < 0.33$, confirmed that the presence of radio halos is associated with merging systems, defined according to X-ray morphology. 

The purpose of the present work is: 
(i) to present and document morphological measurements of the CLASH clusters; 
(ii) to analyze the correlations among their morphological parameters in different spectral bands;  
(iii) to verify whether their morphologies are typical for CLASH-like massive systems in numerical simulations;  
(iv) to quantify the alignment between a cluster's X-ray appearance, its SZE appearance, and its projected mass density as inferred from gravitational lensing; 
and (v) to assess the alignment of Brightest Cluster Galaxies at small radial scales with the larger-scale morphology of the cluster.  
 
\section{The CLASH Project and Sample}

The CLASH cluster program and strategy are described in \citet{2012ApJS..199...25P}. 
Relevant cluster properties are provided in Table 1. 
CLASH was a Hubble Multi-Cycle Treasury program with multiple science goals. 
The most relevant CLASH science goal for this work was to obtain well-constrained gravitational-lensing mass profiles for a sample of 25 massive clusters of galaxies between redshifts of $0.2-0.9$. 
To avoid any biases that would be introduced by selecting clusters on the basis of their lensing signal, twenty of the CLASH clusters were instead selected on the basis of X-ray morphology, to have relatively round X-ray isophotes centered on a prominent BCG.  
The remaining five were selected to be systems capable of providing extraordinary, gravitationally-boosted views of the high redshift universe. 
All of the clusters  have relatively hot intracluster gas (ICM), with global gas temperatures of $kT > 5$ keV. 
This program was allocated 524 orbits over a 3 year (cycle) period between May, 2010 and May, 2013. 
During this time, HST observed all 25 clusters with up to 16 passbands, utilizing the Wide Field Camera 3 (WFC3) Infrared (IR) and UV/Visible (UVIS) channels and the Advanced Camera for Surveys (ACS).  
All CLASH clusters also have good to excellent Chandra X-ray data, with at least 6,000-10,000 X-ray events between 0.5-7.0 keV in publicly available datasets. 

Observations of the CLASH cluster sample have already shown that their concentration-mass relation is consistent with $\Lambda$-CDM based predictions, once the CLASH X-ray morphological selection is taken into account  \citep{2015ApJ...806....4M,meneghetti.etal.2014,2014ApJ...795..163U,2015arXiv150704385U}. 
This major result confirmed suspicions that previous studies, having selected clusters with prominent lensing features, gave mass concentration measurements biased higher than those predicted from simulations based on idealized mass-selection of clusters.  This effect was suggested in \citet[e.g., ][]{2007MNRAS.379..190C,2010A&A...519A..90M,2016MNRAS.455..892G}.
For the CLASH analysis, biases induced by X-ray morphological selection were quantified by selecting clusters from  simulations in the same way that they were selected for inclusion in the CLASH sample. Quantified X-ray morphologies were measured from maps for the simulated clusters with procedures identical to those used on the actual X-ray data \citep{meneghetti.etal.2014}.

In this paper, we provide quantitative X-ray  surface-brightness morphological parameters for the 25 clusters in the CLASH survey \citep{2012ApJ...756..159P}, as well as  similarly-defined morphological parameters derived from gravitational lensing (shear) projected mass maps and SZE gas (Compton ``y"-parameter) maps.
Two sets of morphological measurements were made, one inside a fixed metric aperture of 500 $h_{70}^{-1}$ kpc and
the other inside half the $R_{500}$ overdensity radius 
(i.e. $R_{500}$ is the radius inside which the average mass density is 500 times the critical density at the redshift of the cluster, so it is a cosmology- and mass-dependent quantity.)
We reconstructed the angular scale corresponding to $0.5 R_{500}$  in arcseconds from the $M_{500} h^{-1}$ mass quantity reported in \citet{2015ApJ...806....4M}. 
To avoid ambiguity, we report that specific angular scale for each CLASH cluster in Table~\ref{table:clusters}.
The fixed metric aperture of 500 kpc has the advantage of not changing significantly from analysis to analysis, as
well as having been used by previous observers for similar purposes \citep[e.g.,][]{2005ApJ...624..606J,2010ApJ...721L..82C}. 
For the CLASH sample, a fixed metric aperture is approximately the same fraction of the virial radius for most of the clusters, since these clusters are similar in mass.  
Regardless of the aperture we use, the morphologies of the Brightest Cluster Galaxies (BCGs) at kpc scales in the relaxed sample are strikingly aligned with gravitational potential elongations within these much larger apertures, suggesting a strong relationship between the assembly of the BCG and the cluster as a whole.
 
We also compare CLASH clusters to a broader sample of simulated clusters from MUSIC-2\footnote{MUSIC website: \url{http://music.ft.uam.es}}
 \citep{sembolini.etal.2013}, selected to have similar masses by making a cut at a minimum global temperature of 5 keV, without regard to 
 morphology.
The purpose of this comparison is to examine how typical the morphological properties of CLASH clusters are of massive, simulated clusters in any dynamical state.
Throughout the paper, we assume cosmological parameters of $\Omega_M=0.3$, $\Omega_\Lambda=0.7$, and $H_0=70 h_{70}$ km s$^{-1}$ Mpc$^{-1}$ (i.e. $h_{70}=1$ is the default).

\section{Data Processing}

\subsection{X-ray Imaging}

We use X-ray events from the {\em Chandra X-ray Observatory}, processed and filtered as described in \citealt{2014ApJ...794..136D} (CALDB v4.5.9, CIAO v4.6). 
The data sets are tabulated in Table~\ref{table:clusters}.
Binned X-ray maps were generated with $2 \times 2$ instrument pixel  ($0.984"$) spatial bins with the CIAO script {\em fluximage}. 
Two exposure-corrected images were constructed for each observation ID,
0.7-2.0 keV soft band and a 0.7-8.0 keV broad band. The exposure maps were based on the best bad-pixel maps, aspect solutions
and mask files available, and assumed average energies of 1.0 and 2.0 keV, respectively. Adaptively-smoothed images of the
central $300 \times 300$ kpc from the exposure-corrected 0.7-8.0
keV broad band images are shown in Figure~\ref{figure:X-rayMontage}. The the morphological measurements
were made on images in 0.7-2.0 keV band,  for the most optimal signal-to-noise and to 
minimize the X-ray emission sensitivity to temperature variations.

\begin{deluxetable*}{cccccc}




\tablecaption{CLASH Clusters and Chandra X-ray Observations \label{table:clusters}}

\tablehead{\colhead{Cluster Name} & \colhead{RA} & \colhead{Dec} & \colhead{z} & \colhead{$\rm{0.5R_{500}}$} & \colhead{Chandra} \\ 
\colhead{(---)} & \colhead{(hh:mm:ss)} & \colhead{(dd:mm:ss)} & \colhead{(---)} & \colhead{(arcsec)} & \colhead{Obs ID} } 

\startdata
Abell  209 & 01:31:52.54 &  -13:36:40.4 & 0.206 & 203 & 3579,  522 \\
Abell  383 & 02:48:03.40 &  -03:31:44.9 & 0.187 & 219 & 2321 \\
MACSJ0329-02 & 03:29:41.56 &  -02:11:46.1 & 0.450 & 101 & 6108,  3582 \\
MACSJ0416-24 & 04:16:08.38 &  -24:04:20.8 & 0.396 & \nodata & 10446 \\
MACSJ0429-02 & 04:29:36.05 &  -02:53:06.1 & 0.399 & 113 & 3271 \\
MACSJ0647+70 & 06:47:50.27 &  +70:14:55.0 & 0.584 & \nodata & 3584,  3196 \\
MACSJ0717+37 & 07:17:32.63 &  +37:44:59.7 & 0.548 & \nodata & 4200 \\
MACSJ0744+39 & 07:44:52.82 &  +39:27:26.9 & 0.686 & 74 & 6111 \\
Abell  611 & 08:00:56.82 &  +36:03:23.6 & 0.288 & 149 & 3194 \\
MACSJ1115+01 & 11:15:51.90 &  +01:29:55.1 & 0.355 & 124 & 9375 \\
MACSJ1149+22 & 11:49:35.69 &  +22:23:54.6 & 0.544 & \nodata & 3589,  1656 \\
Abell 1423  & 11:57:17.36 & +33:36:37.47 & 0.213 & \nodata & 11724 \\
MACSJ1206-08 & 12:06:12.09 &  -08:48:04.4 & 0.439 & 109 & 3277 \\
CLJ1226+3332 & 12:26:58.25 &  +33:32:48.6 & 0.890 & 81 & 5014,  3180,  932 \\
MACSJ1311-03 & 13:11:01.80 &  -03:10:39.8 & 0.494 & 81 & 6110,  9381 \\
RXJ1347-1145 & 13:47:30.62 &  -11:45:09.4 & 0.451 & 117 & 3592 \\
MACSJ1423+24 & 14:23:47.88 &  +24:04:42.5 & 0.545 & 82 & 4195 \\
MACSJ1532+30 & 15:32:53.78 &  +30:20:59.4 & 0.363 & 105 & 1665, 1649 \\
MACSJ1720+35 & 17:20:16.78 &  +35:36:26.5 & 0.387 & 115 & 6107 \\
Abell 2261 & 17:22:27.18 &  +32:07:57.3 & 0.224 & 217 & 5007 \\
MACSJ1931-26 & 19:31:49.62 &  -26:34:32.9 & 0.352 & 117 & 3282,  9382 \\
MACSJ2129-07 & 21:29:26.06 &  -07:41:28.8 & 0.570 & \nodata & 3595,  3199 \\
RXJ2129+0005 & 21:29:39.96 &  +00:05:21.2 & 0.234 & 161 & 552,  9370 \\
MS2137-2353 & 21:40:15.17 &  -23:39:40.2 & 0.313 & 148 & 4974,  5250 \\
RXJ2248-4431 & 22:48:43.96 &  -44:31:51.3 & 0.348 & 141 & 4966 \\
\enddata


\tablecomments{ List of CLASH clusters with their RA, Dec, and redshifts from \citet{2012ApJS..199...25P}. The MACSJ0416-24 cluster redshift has been updated since the Postman et al. (2012) work based on VLT and SOAR spectroscopy. The angular quantity $0.5R_{500}$ is computed from lensing-based $M_{500}$ masses derived in  \citet{2015ApJ...806....4M}, in arcseconds, assuming the cosmology assumed in that work. }

\end{deluxetable*}

\begin{figure*}
\includegraphics[width=6.0in]{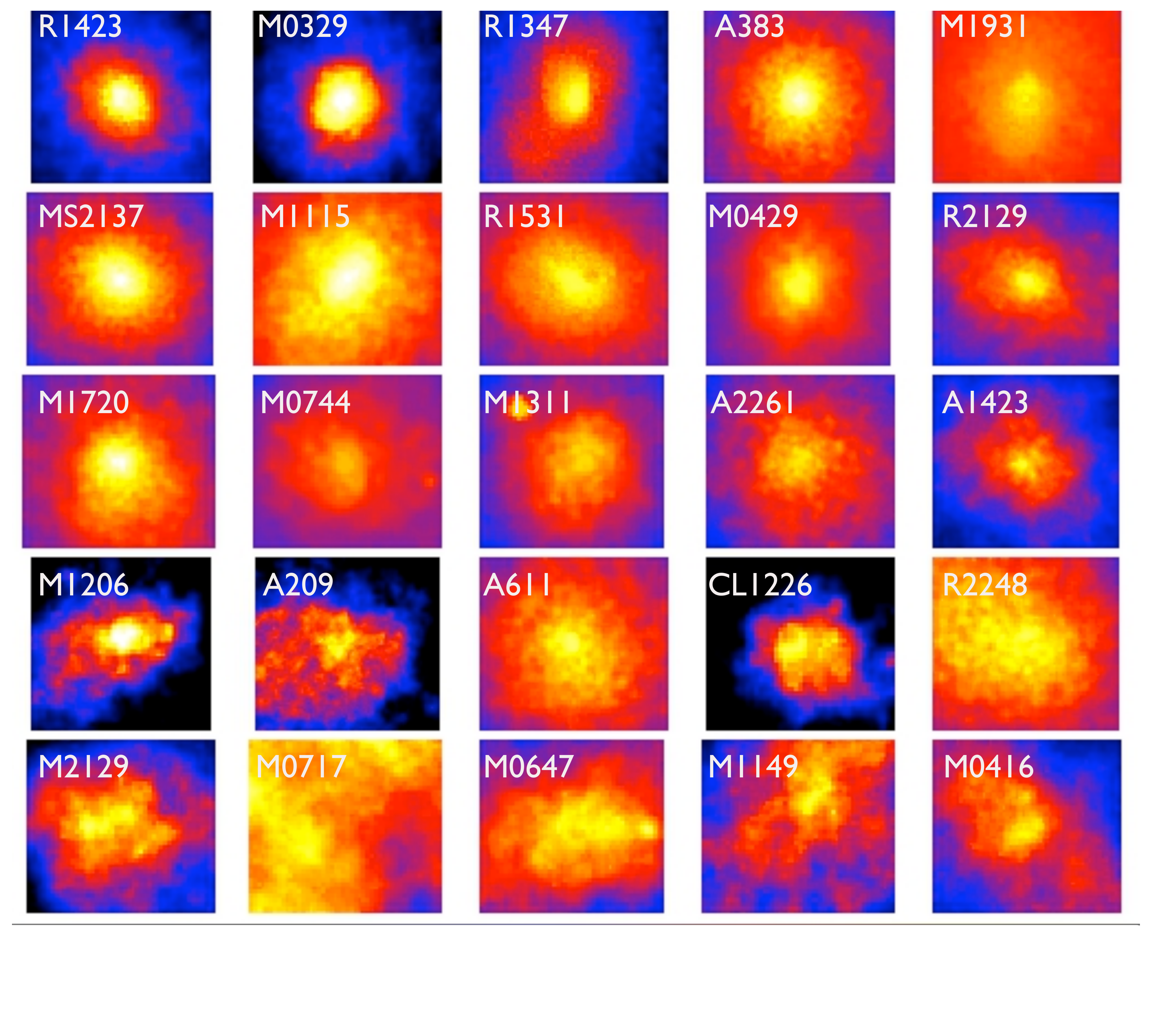}
\caption{\footnotesize
Adaptively smoothed exposure-corrected 0.7-8.0 keV images of all 25 CLASH clusters. This figure shows the central $300 \times 300$ kpc of
each cluster. The clusters are organized from low central gas entropy to the highest central gas entropies, which puts the five
lensing-selected clusters in the last row of 5 panels.
\label{figure:X-rayMontage}}
\end{figure*}

\subsection{Sunyaev-Zeldovich Effect Imaging \label{section:sze}}

The Bolocam Sunyaev-Zeldovich Effect (SZE) images were downloaded from NASA's Infrared Science Archive (IRSA).\footnote{\url{irsa.ipac.caltechEdu/data/Planck/release_2/ancillary-data/bolocam/}}
The details of these data are given in \citet{Sayers2013} and \citet{Czakon2015}.
Specifically, we made use of the data in the file {\it unfiltered\_image.fits}, 
which provides an image of the cluster that are corrected from the
distortion of atmospheric noise filtering, and therefore are 
well suited to constraining morphological parameters.
To characterize the noise in these images, which is correlated
between pixels, we made use of the 1000 bootstrap noise realizations
contained in the file {\it unfiltered\_image\_noise\_realizations.fits}.
The images are $10\arcmin \times 10\arcmin$ in size, and the Bolocam point spread function (PSF)
has a full width half maximum (FWHM) of $58\arcsec$. The SZE brightness and noise varies over the 
CLASH sample, and the peak S/N per resolution element in the images ranges from 5 to 40.

\subsection{Lensing Models and Maps \label{section:lensmodel}}

Gravitational lensing maps of the surface mass density ($\kappa$) have been constructed utilizing both
strong- and weak-lensing information from the Hubble Space Telescope. The analysis is described more fully in \citet{2015ApJ...801...44Z}, and here we give a very brief summary. The lens modeling was performed using two complementary parametric methods, to obtain a better grasp on systematics. The light-traces-mass (LTM) method assumes that light traces mass for both the galaxies and the dark matter, with the latter being a smooth version of the former, and the two components are added with a free relative weight. The second method (PIEDeNFW) assumes LTM for the cluster galaxies but then fits an analytical elliptical NFW form \citep{1997ApJ...490..493N} for the dark matter  (PIEMDeNFW: Pseudo-Isothermal  Elliptical  Mass  Distribution plus elliptical Navarro Frenk \& White profile). 
The minimization is performed via a Monte Carlo Markov Chain. The lensing maps of the best fit model are then generated on a grid similar to the original HST image used to define the input,  with a spatial resolution of 65 milliarsceconds per pixel. Statistical uncertainties were estimated from 100 random steps from the MC chain.
We use the second method (PIEMDeNFW) for the lensing map for our baseline analysis, and we estimate systematic uncertainties by comparing our baseline result to an identical analysis using the LTM method.

Two interesting findings from \citet{2015ApJ...801...44Z} are worth noting here. 
The first is that systematic uncertainties dominate the lensing error budget and are about $40\%$ in 
$\kappa$, on average, per pixel, among all CLASH clusters. 
\citet{2015ApJ...801...44Z} determine that typical errors on lensing quantities are thus underestimated,  due to traditional use of only one method per cluster.  Efforts have been made in the past couple of years to overcome this and learn about systematics in lens- and strong-lens modeling in particular. However, various factors of uncertainty such as errors from large scale structure or other correlated matter along the line of sight \citep[e.g., ][]{2012MNRAS.420L..18H,2011MNRAS.411.1628D} are sometimes not accounted for properly in error budgets. 
\citet{2015ApJ...801...44Z} therefore recommend, in using these maps, to estimate the typical systematic error from the 
differences between the two methods they employ. 
These systematic differences in $\kappa$ do not lead to significant variance in our scientific conclusions.

\subsection{Simulations and Mock Cluster Catalogs}

The MUSIC-2 sample includes 282 clusters selected within a
cosmological dark-matter box of volume (1 $h^{-1}$ Gpc)$^3$ and
re-simulated with a hydro component at higher resolution
\citep{sembolini.etal.2013}.  The parent simulation (the {\em
  MultiDark} simulation) was carried out with the code ART
\citep{kravtsov.etal.97} and contained 2048$^3$ particles. The
underlying cosmological model is identical to what we have
assumed in this paper. They assume a flat universe described by the
following values of the cosmological parameters: $\Omega_M=0.3$ as matter density;
$\Omega_b=0.0469$ as baryon density; $\sigma_8=0.82$ as primordial
amplitude of fluctuation in a scale of 8 $h^{-1}$ Mpc; $n=0.95$ as
power spectrum index; and $h=0.7$ as reduced Hubble parameter ($h_{70}=1$).
The CLASH sample is comprised of clusters with gas temperatures above 5 keV. 
Assuming $M-T$ relations from \citet{2009ApJ...692.1033V}, this temperature limit corresponds to the 
a mass limit equal to $M_{500} > 3.5 \times10^{14} ~\rm{M}_{\odot} h^{-1}$, which 
is satisfied by $\sim 100$ clusters at $z=0.333$ 
(only 1 simulated cluster has mass $M_{500}> 10^{15}~\rm{M}_{\odot} h^{-1}$ at that redshift).

The re-simulations involved Lagrangian regions of 6 $h^{-1}$ Mpc
radius around the most massive halos (with virial masses above
$10^{15} h^{-1} M_{\odot}$ at redshift $z=0$) and were performed by using the TREEPM+SPH code GADGET
\citep{springel2005}. Two sets of re-simulations were produced including both
non-radiative and radiative physics. In this work, as in
\citet{meneghetti.etal.2014}, we focus on the non-radiative simulation
because the radiative run did not include any prescription for feedback by
active galactic nuclei, implying that the cluster core is affected by over-cooling \citep{borgani&kravtsov}.
 In the radiative case, the condensation of X-ray luminous gas in the center is extreme
\citep{rasia.etal.2013conc}, the light concentration is not realistic, and the cluster isophotal shapes are
less in agreement with observations than non-radiative simulations \citep{lau.etal.2012}. 
 We note that in a comparison project of various codes and AGN-feedback implementation schemes \citep{2015arXiv151103731S},  
that most of the simulations that include AGN feedback  give  similar dark matter distributions and gas fractions as those found  
in the non-radiative simulations outside of cluster cores.  

The mass resolution of the MUSIC-2 simulations corresponds to $m_{\rm
  DM}=9.01 \times 10^8 h^{-1}M_{\odot}$ for the dark-matter particles
and $m_{\rm gas}=1.9 \times 10^8 h^{-1} M_{\odot}$ for the gas
particles. The gravitational softening was set to $6 h^{-1}$ kpc.  Ten different simulation
snapshots were stored. We analyze those four that cover the same redshift range of the CLASH sample:
$z=0.25, 0.33, 0.43, 0.67$, respectively.

\subsubsection{Mock X-ray Catalog}
For each simulated halo, we produce three {\it Chandra}-like event files
corresponding to the orthogonal line-of-sight projections of the original cosmological volume. 
These projections are therefore randomly oriented relative to a cluster. 
The three files are not co-added and are analyzed independently.  
The tool adopted is the
X-ray MAp Simulator (X-MAS,\citealt{gardini.etal.2004,rasia.etal.2008}) 
which accounts for the ancillary response function and the redistribution matrix function 
of the ACIS-S3 detector on board of the {\it Chandra} satellite. 
The field of view is set equal to 16 arcmin, which corresponds to $\sim 4.5$
 Mpc at $z=0.333$. The two outer radii used in the current analysis, 500 kpc and $0.5 R_{500}$,
are amply within the mock X-ray image. To generate the
event files we assume a fixed metallicity with value equal to 0.3
solar (as tabulated by \citealt{1989GeCoA..53..197A}); we include the
galactic absorption with a WABS model ($N_H=5\times 10^{20}$
cm$^{-2}$); and we impose an exposure time equal to 100 ks.

As we have done for the real X-ray observations, to evaluate the morphological
parameters from the mock catalogue, we produce soft X-ray band ([0.5-2]
keV) images binned by $2\times 2$ arcsec$^2$. 
For a detailed description of the method, see \citealt{rasia.etal.2013}.

\subsubsection{Mock SZE catalogue}
To evaluate the effect of the Bolocam PSF on the morphological parameters 
(see Section 4.2), we produce maps of the Compton $y$ parameter. 
For each simulated cluster, we chose only one line of sight, centered the map
on the minimum of the potential well, and similarly to the X-ray images consider 
a field of view of 16 arcmin as a side and an integrating distance of 10 Mpc. 
The resolution of each pixel is 1 arcsec.

\section{Morphology Measurements}

\subsection{Morphologies \label{section:defs}}

For the following discussion we define general quantities based on analysis of maps of
scalar observables. 
In the discussion to follow, we will talk in terms of the
surface brightness of light. However, we also map the SZE Compton y-parameter or
surface mass density for a cluster and characterize the distribution using
identical conventions applied to maps of the light distribution.
In this study, we quantify 2-D cluster morphology using the following parameters:

\begin{enumerate}

\item {\it Concentration, $C$} is defined here to be the ratio between the light (or other mapped 
observable) within a circular aperture with a radius $R_{inner}$ and the
total light enclosed within a circular aperture with a radius $R_{outer}$. The concentration $C$ is defined where $(R_{inner},R_{outer}) = 
(100, 500)$ kpc \citep[See also ][]{2010ApJ...721L..82C}. For the case where we use scaled apertures to define the radii, $(R_{inner},R_{outer}) = 
(0.15, 0.5) R_{500}$. Note that this $C$ is {\em not} the same concentration $c$ used in the mass-concentration relation, nor is
it based on the percentage of total enclosed light, as is occasionally used elsewhere for galaxies and X-ray clusters \citep[e.g.,][]{1994ApJ...432...75A}. 

\item {\it Centroid shift, $w$} \citep[e.g.][]{2006MNRAS.373..881P, 2006ApJ...639...64O, 2008ApJ...685..118V, 2008ApJS..174..117M, 
2010A&A...514A..32B, 2010ApJ...721L..82C,rasia.etal.2013} is the standard deviation of the projected separation between the X-ray peak 
and centroids estimated within circular apertures of increasing radius up to $R_{\rm max} = 500 h_{70}^{-1}$ kpc:
\begin{equation}
w = \left[ \frac{1}{N-1} \sum_i \left( \Delta_i - \bar{\Delta} \right)^2 \right]^{1/2} \frac{1}{R_{\rm max}}
\end{equation}

\item {\it Power ratio, $P_3/P_0 \equiv P30$,$P_4/P_0 \equiv P40$ }: the power ratios are defined from the multipole decomposition of the 
two-dimensional X-ray surface brightness in circular apertures centered on the cluster's centroid:
\begin{equation}
\frac{P_m}{P_0} = \frac{(a_m^2+b_m^2)/(2 m^2 R^{2m})}{\left[ a_0 \ln( R ) \right]^2}
\end{equation}
where $a_m( R ) = \int_{R<R_{\rm max}}{S({\bf x}) R^m cos(m\phi) d^2x}$, $b_m( R ) = \int_{R<R_{\rm max}}{S({\bf x}) R^m sin(m\phi) d^2x}$, 
${\bf x} = (R, \phi)$, $S({\bf x})$ is the X-ray surface brightness at sky location ${\bf x}$. The ratio $P_3/P_0$, estimated from the above equations with $m=3$, 
is sensitive to deviations from mirror symmetry and insensitive to ellipticity, in the sense that a high value of $P_3/P_0$ 
indicates a clear asymmetric structure 
in the ICM (See e.g. \citealt[][]{1995ApJ...452..522B, 1996ApJ...458...27B, 2005ApJ...624..606J, 2008ApJ...685..118V,2010A&A...514A..32B, 2010ApJ...721L..82C,
rasia.etal.2013}.) Power ratios for clusters of galaxies are typically quite tiny ($\sim 10^{-7}$) with statistical uncertainties not much smaller, but they span a significant dynamic range (4 orders of magnitude in the case of CLASH clusters).

\item {\it Axial ratio, (AR)} is a measure of the elongation of the cluster surface-brightness map. We use the same 
procedure as documented in \citep{2015ApJ...805..177D}.
Briefly, the axial ratio is estimated 
from the ratio of the non-zero elements after the diagonalization of a symmetric $2\times2$ matrix with elements equal to the second moments 
of the surface brightness, computed at each pixel element $(x,y)$ within the given aperture centered on the centroid. 
AR is defined as the ratio between the lengths of the major and minor axes, 
with values between 0.0 and 1.0 ($AR=1.0$ corresponds to the circular case). 

\item {\it Position angle, (PA)}, by astronomical convention, 
is the alignment of the semi-major axis in degrees East of North. It is computed simultaneously with the AR \citep{2015ApJ...805..177D}.
Briefly, it is the rotation angle in degrees required to bring the matrix of the second moments into
its diagonal form, and we correct this formal angle to the astronomical convention. 
\footnote{ The IDL routine ELLFIT, which uses the same conventions described here, 
was used to derive eccentricity/ axial ratio and the position angle. 
\url{http://www.astro.washingtonEdu/docs/idl/cgi-bin/getpro/library09.html?ELLFIT}.  }
Because of the degeneracy of rotation of the long axis, we occasionally add or subtract 180 degrees in plots in order to
more easily compare PA measurements made for the same cluster, but from different maps.
\end{enumerate}

\subsection{X-ray Morphologies \label{section:Xraymorph}}

We estimated uncertainties in the morphological parameters by measuring multiple versions of the maps.
The error budget for each X-ray morphological parameter was estimated by Monte-Carlo methods, 
by re-sampling the counts per pixel according to their Poissonian error to make statistically similar maps as 
in \citep{2010ApJ...721L..82C}. 

The X-ray results are summarized and reported in Table~\ref{table:xraymeans}. 
X-ray results for individual clusters are
reported in Table~\ref{table:xray500morph}.

For all comparisons we make between the X-ray morphological parameters and the same parameters derived from the SZE, lensing,
and simulated maps,
we have chosen a fixed metric radius of 500 kpc. Ideally, we would choose a fixed fraction of a radius derived by
a mass overdensity, such as half of the $R_{500}$ radius, which is defined to be the radius inside which the mean
density is 500 times the critical density at the same redshift. However in practice, this radius can be difficult to 
work with in an analysis, since it is dependent on the mass estimate, and there are multiple technique-dependent mass measurements for any given cluster.
For the CLASH clusters, $0.5 R_{500} $ is approximately 500 kpc for each cluster. A 500 kpc radius turns out to define an area sampled
by all of the X-ray and SZE measurements without significant extrapolation. So for comparison of measurements of all 25 
CLASH clusters from one technique to another, we decided that the optimal choice was 500 kpc. For the clusters where
we have measurements at both scales, with $R_{500}$ defined by the lensing measurements of \citet{2015ApJ...806....4M},
we show in Figure~\ref{figure:comp500} that measurements made at slightly different radii, unsurprisingly, are
strongly correlated with each other. Except for concentration, 
the scatter in each case is computed relative to an identity line, not a fit. 
For concentration and centroid shift, the measurement errors are smaller than the scatter, but the typical variation is small in
both cases, about 3\% and 0.2\% respectively. The dispersion in position angle is about 6 degrees while the 
scatter in difference of the axis ratios is about 2\%, considerably less than the measurement uncertainty.

There is significant intrinsic scatter in the measurements in concentration, and  this real variance is due to the 
direct dependence of the concentration parameter on how the outer radius is defined. Therefore when we compare concentration
measured from X-ray maps to that measured on other maps, using the exact same inner and outer radius for the concentration ratio 
is important. The concentration defined at $0.5 R_{500}$ is strongly correlated with that computed at 500 kpc,
such that a best linear fit to the measurements, including errors in x and y, 
yields $C( \rm{at~0.5R_{500}}) = 0.95 (\pm 0.01) C( \rm{at~500~kpc}) +0.078 (\pm0.006)$. We plot this relation along with the line representing $C( \rm{at~0.5R_{500}}) = C( \rm{at~500~kpc})$ in Figure~\ref{figure:comp500}. 
Figure~\ref{figure:comp500} shows that the other parameters, centroid $w$, position angles, and axis ratios, can be measured in apertures of slightly different sizes without changing the nature of the correlation.
We note that whenever we compare measurements from one technique to another for an individual cluster, we make that comparison over the identical region on the sky for each measurement.

In Table~\ref{table:xraymeans}, we compare the mean and dispersion for morphological properties measured within 500 kpc
and within $0.5 R_{500}$ for the X-ray selected subsample, as well as measurements made within 500 kpc for the full
sample. We note that the means and dispersions for properties measured for the X-ray selected CLASH subsample
are virtually identical to each other, whether measured in the 500 kpc aperture or within $0.5 R_{500}$. 
 To test for radial variations, we repeated our morphological analysis at outer radii of 100, 200, 300, and 700 kpc for our X-ray maps, and at
200, 300, and 700 kpc for the lensing maps. We found no significant radial trends in measurements of axial ratios
or position angles. The axial ratios remain within about 2\% for all radii except for the innermost X-ray radius at 100 kpc,
where the axial ratio is about 5\% rounder than at 500 kpc. The averaged absolute 
measured position angles vary very little with radius (in the X-ray maps, the average difference was 
$\lesssim 10-15$ degrees at radii between 200-700 kpc, and for the lensing maps the difference was $\lesssim 4$ degrees.)

Therefore for this study we are comfortable with using the 500 kpc aperture throughout, which is the only aperture common to all
the clusters in the full range of measurement techniques considered in this paper. Note that the medians 
and standard deviations for the same quantity is similar across all samples except for the ``SL" (lensing-selected) sample, which 
is constituted of the most obviously non-relaxed systems in the CLASH sample. We summarize the comparison here:   
the X-ray images of the X-ray-selected CLASH sample are
more centrally concentrated ($\sim 0.4$ vs $\sim 0.1$), have smaller centroid shifts ($w \sim 0.005$ vs $w \sim 0.02$),
more circular (AR $\sim 0.9$ vs $\sim 0.8$), and have higher moment ratios $P30$ and $P40$ about an order of magnitude
smaller than X-ray images of clusters in the lensing-selected sample.

 The concentration $C$ has been used as an inexpensive surrogate for identifying candidate cool core clusters 
\citep[e.g., ][]{2008A&A...483...35S,2010A&A...521A..64S,2012ApJ...761..183S}. Cool core clusters tend to be more concentrated than non-cool core clusters, and this trend applies to the CLASH sample as well.
The radial bounds for the definition of $C$ in this work 
differ somewhat from the definitions used for the studies by Santos and Semler ($R_{inner}-R_{outer}$ are 
$40-400$ kpc h$_{70}^{-1}$ instead of $100-500$ kpc h$_{70}^{-1}$, 
but the chosen energy range is very similar to that used by \citet{2010A&A...521A..64S} (we use a lower bound of 0.7 keV rather than 0.5 keV). 
If we classify a cool-core cluster as having a central gas entropy $K_0$ of $kT n_e^{-2/3} \sim 30$ keV cm$^{2}$ \citep[e.g.][]{2009ApJS..182...12C}, the 
approximate threshold between cool-core clusters and non-cool core clusters in CLASH is $C \sim 0.4$, defined using the 
central gas entropies reported by
\citet{2015ApJ...805..177D}. (The result is insensitive to whether we define cool cores by their central gas entropy or their cooling times.) 
There are 11 clusters in CLASH with $K_0<30$ kev cm$^{2}$. All eleven have $C<0.4$. Only one cluster has a higher central entropy and similar surface brightness concentration, MACSJ1311-03, with $K_0=47\pm6$ keV cm$^{2}$ and $C=0.49\pm0.02$.  
We see no correlation between $C$ and $K_0$ for the low $K_0$ systems. 

\begin{deluxetable}{lccc}
\tablecaption{Median X-ray Morphologies for the CLASH Samples \label{table:xraymeans}}
\tablehead{\colhead{Quantity} & \colhead{Sample} & \colhead{Mean} & \colhead{N} }
\startdata
$C$ 500kpc & X-ray          & $0.43\pm0.13$ &            20 \\
$C$ r500/2 & X-ray*         & $0.48\pm0.12$ &            19 \\
$C$ 500kpc & all          & $0.37\pm0.16$ &        25 \\
$C$ 500kpc & SL      & $0.11\pm0.08$ &             5 \\
$w$ 500kpc & X-ray         & $0.005\pm0.010$ &            20 \\
$w$ r500/2 & X-ray*             & $0.005\pm0.008$ &            19 \\
$w$ 500kpc & all          & $0.006\pm0.012$ &        25 \\
$w$ 500kpc & SL      & $0.020\pm0.010$ &             5 \\
Log P30 500kpc & X-ray       & $-7.03 \pm 0.46$ &            20 \\
Log P30 r500/2 & X-ray*      & $-7.30 \pm 0.60$ &            19 \\
Log P30 500kpc & all         & $-6.90 \pm 0.66$ &        25 \\
Log P30 500kpc & SL    &     $-5.80 \pm 0.40$ &             5 \\
Log P40 500kpc & X-ray       & $-7.45 \pm 0.46 $ &            20 \\
Log P40 r500/2 & X-ray*           & $-7.62 \pm 0.52$ &            19 \\
Log P40 500kpc & all        & $-7.35 \pm 0.63$ &        25 \\
Log P40 500kpc & SL    & $-6.25 \pm 0.56$ &             5 \\
Axial Ratio 500kpc & X-ray    & $0.91\pm0.05        $ &    20 \\
Axial Ratio r500/2 & X-ray*   & $0.89\pm0.05        $ &    19 \\
Axial Ratio 500kpc & all & $0.90\pm0.06       $ & 25 \\
Axial Ratio 500kpc & SL & $0.81\pm0.06       $ &      5 \\
PA Difference (Deg)   & X-ray* & $0.9\pm 11         $ &    19 \\
\enddata
\tablecomments{ Medians and standard deviations of the X-ray based morphological quantities for subsamples of the CLASH clusters. The ``all" sample is all 25 CLASH clusters. The ``X-ray" sample is all 20 X-ray selected clusters measured at 500 kpc. The ``X-ray*" sample is the 19 X-ray selected clusters with measurements out to $0.5 R_{500}$. The ``SL" sample is the 5 lensing-selected CLASH clusters. The PA difference reported in this table is the difference in degrees between the orientation of the longest axis, measured at 500 kpc vs. $0.5 R_{500}$, showing that the exact definition of the aperture does not affect the PA estimate.}


\end{deluxetable}

\begin{figure*}
\includegraphics[width=6.0in]{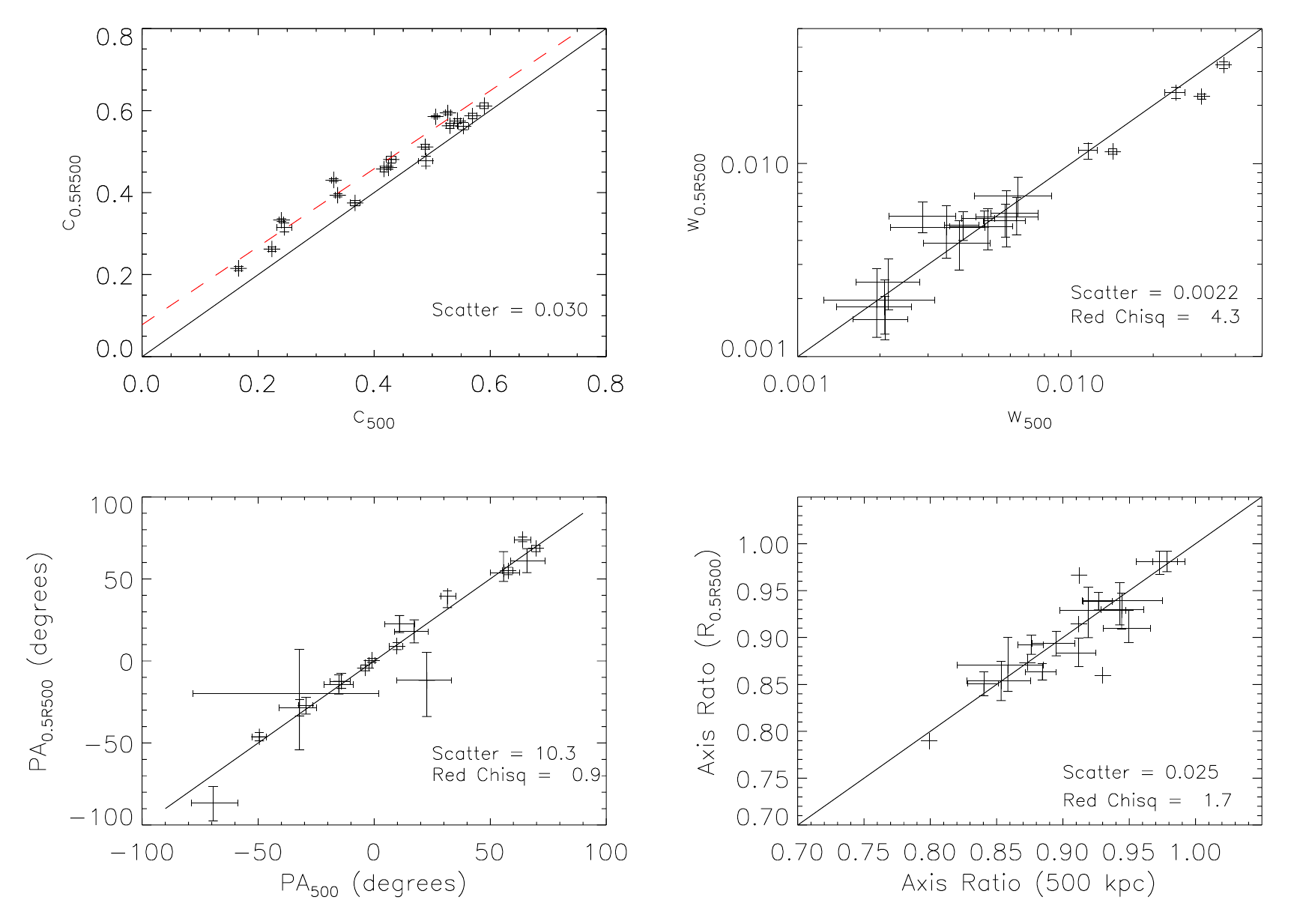}
\caption{\footnotesize
 Comparison of X-ray measurements at $500$ $h_{70}^{-1}$ kpc and at $0.5 R_{500}$, for the 19/25 (19/20 X-ray selected) CLASH clusters for which direct comparison is possible. Solid black lines representing the equality line is overplotted for each. The scatter and reduced $\chi^2$ (Red Chisq) is in reference to that line for $w$, PA, and Axis Ratio. The best fit to the concentration measured at 500 kpc vs. that at $0.5 R_{500}$ is plotted in a dashed red line in the upper left panel, and discussed in the text.
  \label{figure:comp500}
  }
\end{figure*}

\subsection{Lensing Morphological Parameter Estimates \label{section:Lensmorph}}

Using the lensing maps, we computed the  centroid shift, axial ratio and position angle.
We did not compute power ratios for the lensing data since the uncertainties did not yield interesting results.
We utilized otherwise identical procedures for quantifying the morphological parameters for the 
lensing maps. 

As discussed in \S~\ref{section:lensmodel}, the uncertainties for morphological parameters derived from lensing
maps was estimated based on the differences between the morphology measurements made from the two different techniques. We verified that 
those systematic uncertainties are larger than the statistical estimates obtained from re-measuring 100 statistically re-sampled maps, 
and reflect a better quantification of the uncertainties of the estimates.
Our results for deriving morphological parameters from the lensing-based surface mass density maps for individual clusters are reported in Table~\ref{table:lens500morph}. 

\subsection{SZE Morphological Parameter Estimates \label{section:SZEmorph}}

We computed the concentration, centroid shift, axial ratio, and position angle of the 
SZE images using the same procedures listed in Section~\ref{section:defs}. However, due to the limited
spatial dynamic range of the SZE images, along with their modest S/N, we did not
compute the value of the power ratios. In all cases, the same center positions and 
outer radii used for the X-ray and lensing analyses were also used for the SZE
analysis.

The relatively large size of the Bolocam PSF produces a bias in some of the derived morphological
parameters, particularly the value of the concentration and centroid shift.
We correct for this bias using mock SZE observations of the simulated clusters
from the MUSIC-2 sample according to the following procedure.
First, we compute the true values of the morphological parameters using the
mock SZE observations at the native resolution of the simulation.
Next, the mock SZE observations are convolved with the Bolocam PSF, and the
morphological parameters are recomputed.
We then perform a linear fit to the true parameter values derived from the 
unconvolved mock observations and the (in general) biased parameter values
derived from the PSF-convolved mock observations.
A separate linear fit is performed for the clusters within each of the four
redshifts of the MUSIC-2 sample (0.250, 0.333, 0.429, and 0.667).
These linear fits, interpolated to the redshift of each real cluster in the 
CLASH sample, are then used to correct for the PSF-induced bias in the 
morphological parameters derived from the Bolocam data.
In addition, the scatter in the mock-observation-derived values
relative to the linear fits is added as a systematic
uncertainty to all of the SZE results.

The SZE images contain noise that is correlated among pixels, and  
noise on large angular scales produces features that mimic deviations
from spherical symmetry. In order to correct for this noise-induced
bias we compute the value of the morphological parameters for each
of the 1000 bootstrap noise images. In the case of the centroid shift
and the axial ratio, the mean value determined from these noise fits
is significantly different from the nominal values of 0 and 1. 
Therefore, for these two parameters we correct the value derived
from the actual Bolocam images according to the mean value
derived from the bootstrap fits. In some cases, due to noise fluctuations,
this correction results in a best-fit parameter value outside
of the physically allowed region. For example, a centroid shift that 
is less than, but statistically consistent with, zero.

The SZE-derived morphological properties within 500~kpc apertures
are listed in Table~\ref{table:SZ500morph}.

\section{Discussion}

\subsection{X-ray Morphology Correlations}

In general, we would expect that clusters with smaller X-ray concentration ($C$), larger centroid shift ($w$), and larger
power ratios would be more likely to be disturbed clusters, a trend that can be seen in Table~\ref{table:xraymeans} and summarized in \S~\ref{section:Xraymorph}. 
We also might expect these measures to be loosely correlated with each other.  Inspection of the CLASH morphologies
plotted in Figure~\ref{figure:morph} shows the expected qualitative correlations seen in the $w-C$ diagram by
 \citet{2010ApJ...721L..82C} in their sample. The quadrants in this figure were defined by \citet{2010ApJ...721L..82C}.
The X-ray selected CLASH clusters are mostly relaxed in Figure~\ref{figure:morph}. 
Fourteen of the 20 X-ray selected clusters lie in the upper-left quadrant, and none are located in the lower-right quadrant, 
which is where the most-disturbed clusters are. 


\begin{figure}
  \begin{minipage}[b]{0.5\linewidth}
\includegraphics[width=3.5in]{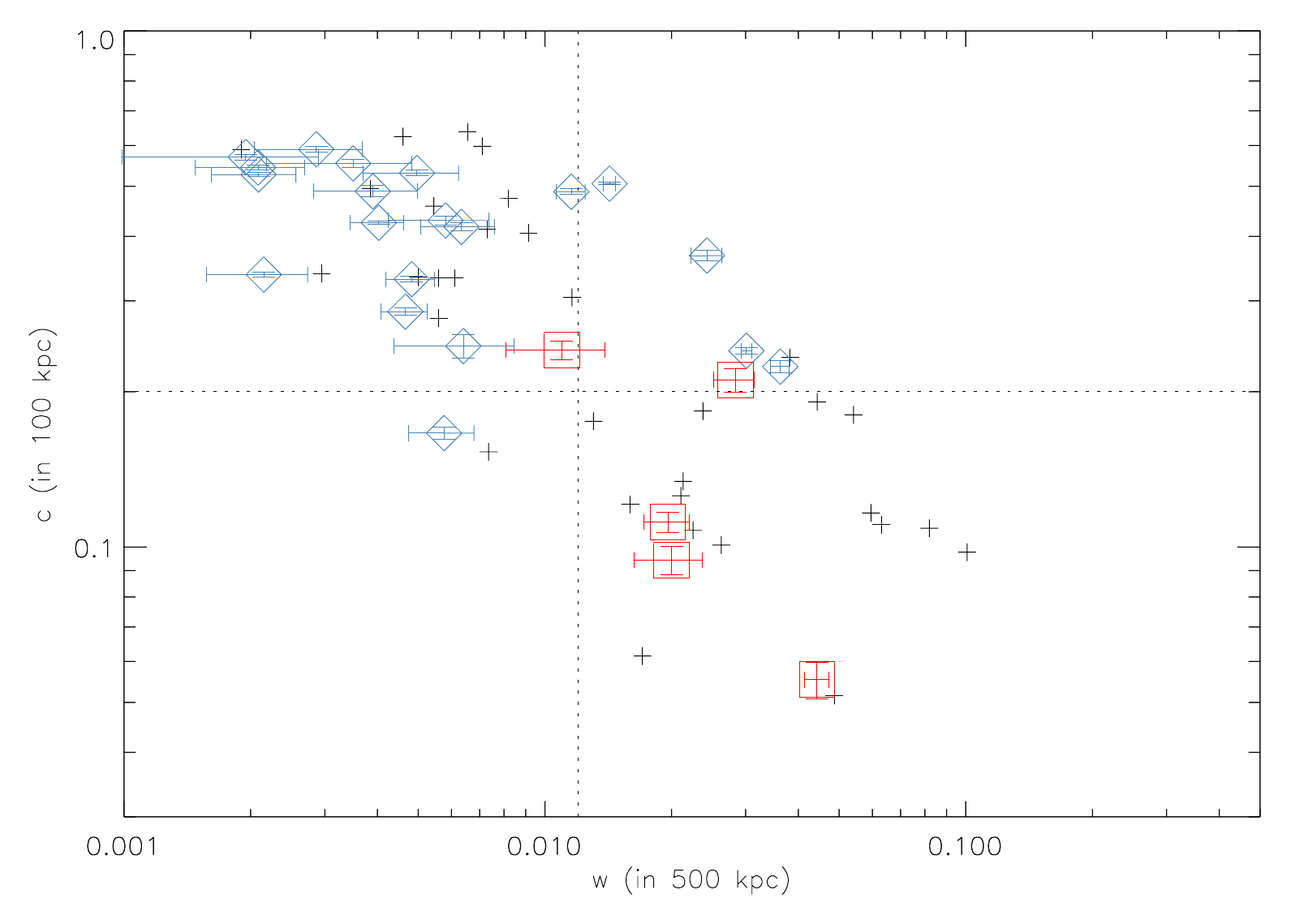}
  \end{minipage}
\caption{\footnotesize
  X-ray concentration is inversely correlated with centroid shift. 
  Crosses are points from Cassano et al 2010 (C10). Blue diamonds/red squares are for CLASH X-ray selected/lensing-selected clusters.
  Dashed lines are lifted approximately from C10. The lower right hand quadrant marks out the more disturbed clusters, while
  the upper left hand quadrant is occupied by more relaxed clusters.
  \label{figure:morph}
  }
\end{figure}

We see that the X-ray morphologies of CLASH clusters show similar correlation between morphological parameters as
seen in \citet{2010ApJ...721L..82C}. The CLASH clusters that were lensing-selected have morphological indications that they are
disturbed systems, similar to clusters in the Cassano sample that showed evidence for merger activity from
the presence of radio halos and X-ray indicators.

\subsection{Comparison between X-ray and lensing and SZE morphologies}

We have compared the morphological parameters from X-ray and lensing maps in Figure~\ref{figure:xraylens}
and the SZE maps in Figure~\ref{figure:xraysz}. Only the morphological property of position angle 
 correlates across all of the map classes.  The lensing-X-ray PA correlation has
only one distinct outlier, MACS1206. The discrepancy for this cluster disappears if its X-ray PA is measured
at slightly smaller or larger radii. For all other clusters, the X-ray (or lensing) PA is not sensitive to choice of 
measurement radius for radii larger than 200 kpc.  The other apparent outliers have large uncertainties. 
MACS0744 has a very uncertain X-ray position angle,
and MACS0717, a highly irregular system, has large systematic uncertainties in the determination
of the orientation in its lensing map (as do CL1226 and Abell 1423, to a somewhat lesser extent.)
 But for the majority of the CLASH clusters, the lensing PA at 500 kpc is quite similar to the one inferred from the X-rays. 
  
The situation is much noisier in the SZ-X-ray comparison, due mainly to the larger PSF and lower S/N in the SZE maps. There is a clear correlation between the
SZE and X-ray PA values, although there are also several statistically significant outliers. The cause of these outliers may be physical in nature, for example due
to a difference in the gas orientation between the central region where most of the X-ray signal originates and the outer regions near 500 kpc where a large
fraction of the SZE signal originates. Further, the presence and locations of shocks and/or high pressure regions could also produce differences in the SZE and
X-ray measured PA values. However, the outliers may also be a result of systematics related to the SZE analysis. Although we have developed a rigorous
procedure to correct for the large PSF and the large angular scale noise in the SZE images, either or both of these effects could potentially bias our derived PA
values.

Two effects could cause the dynamic range of the X-ray measurements of concentration and axis ratios to exceed
those of the same properties measured from lensing and SZE maps.
\begin{enumerate}
\item The Chandra X-ray maps have an instrumental PSF with a width $\sim1\arcsec$. The effective PSF is photon limited
to be larger, however, the effective Chandra PSF for emissivity fluctuations is considerably more compact than the effective
resolution of large scale structures in the lensing and SZE maps. 
\item X-ray surface brightness of the
X-rays scales like density squared, 
as opposed to linearly in gas density (or pressure) 
for the SZE signal and linearly in projected total mass density for the lensing signal. As a result, structures of higher density, such as the
central region of the cluster, will produce an enhanced X-ray signal compared to the SZE or lensing signal.
\end{enumerate}

We find the typical axis ratio for CLASH clusters in the X-rays to be $0.88\pm0.06$, which is 
similar to the SZE maps at $0.90\pm0.06$, and somewhat
more elongated (at 500 kpc) in lensing maps $0.80\pm0.08$ (although the LTM lensing models
are more circular, at $0.92 \pm 0.04$.) In a one-to-one comparison, the lensing
maps are more elongated than the projected X-ray emission, but are generally aligned in the same direction.
That these clusters are typically circular is
not surprising, since they were selected to be nearly circular in the X-ray. That they are similarly circular in SZ
and lensing images then is also expected. That the gas is about 10\% or so rounder than the
projected mass at 500 kpc was predicted by \citet[e.g.][]{2007MNRAS.377..883F}  
when the total mass is dominated by collisionless dark matter, and thus the relative axial ratios (and therefore eccentricities)
is consistent with gravitational potentials dominated by collisionless dark matter.

The SZE estimates of the axis ratios are not particularly correlated with the X-ray estimates
at the same radius, but both estimates have a very similar mean and standard deviation, $0.9 \pm 0.06$, where the
scatter is dominated by the measurement uncertainties.
Abell 1423's SZE axis ratio is an outlier for the sample's range of SZE measurements, 
possibly due to the dim SZE signal towards this cluster. The two lowest X-ray axis ratios (i.e the highest elongations) 
were found for MACS0416 and MACS0647.   These are two of the five lensing-selected clusters from the CLASH survey \citep{2012ApJS..199...25P}.
Both of these CLASH clusters have evidence for interactions in their optical appearance (at least two bright
galaxies in their core, with extended elongated intracluster light in between them.)

We have plotted the histograms of centroid and position angle differences in Figure ~\ref{figure:hist_cent}
and ~\ref{figure:hist_PA}.

\begin{figure*}
\includegraphics[width=6.0in]{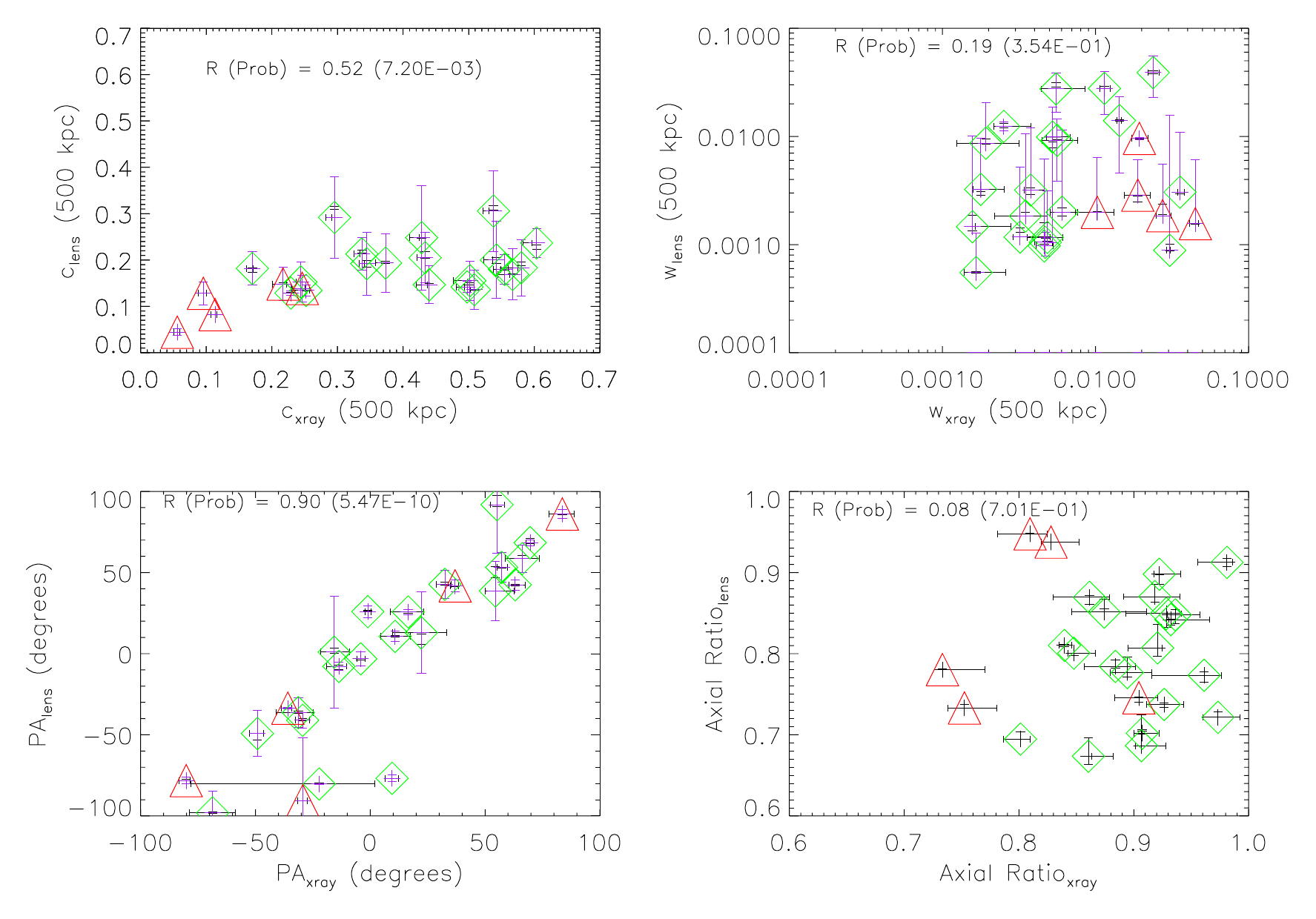}
\caption{\footnotesize
   Morphological properties measured with $r_{outer}=500 h_{70}^{-1}$ kpc for X-ray and lensing maps.
   The X-ray selected subsample is plotted with green diamonds. The lensing-selected subsample
   is plotted with red triangles. The black error bars are statistical. The systematic uncertainty for
   the lensing estimates are represented by the purple error bars. These uncertainties are based on the
   difference in results between the analyses of lensing maps from two different methods in Zitrin et al. 2015. 
   Spearman's rank coefficient and probability (lower probabilities are more significant) were computed 
   and reported in each plot for the full sample.  
  \label{figure:xraylens}
  }
\end{figure*}

\begin{figure*}
  \begin{minipage}[b]{0.5\linewidth}
\includegraphics[width=6.0in]{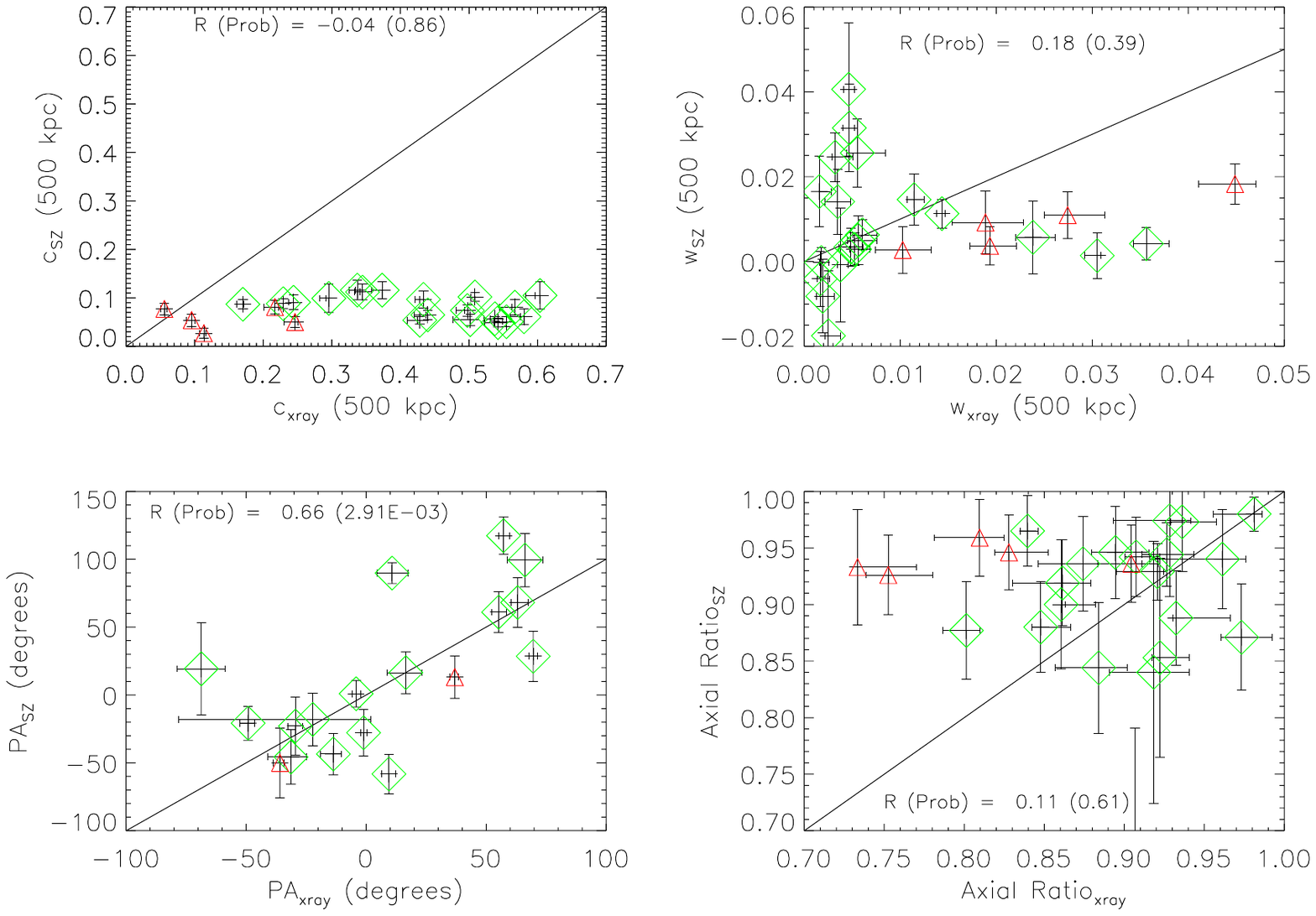}
  \end{minipage}
\caption{\footnotesize
   Morphological properties measured with $r_{outer}=500 h_{70}^{-1}$ kpc for X-ray and SZE maps.
The error bars for the SZE morphology are based 
  on bootstrapped SZE maps as described in the text. Negative values for SZE $w$ arise from the correction prescription
  for the large angular scale noise, as described in the text. The X-ray selected subsample is plotted with green diamonds. The lensing-selected subsample   is plotted with red triangles.  Spearman's rank coefficient and probability (lower probabilities are more significant) were computed 
   and reported in each plot for the full sample.  The PA is either unconstrained or poorly constrained in several of the SZE maps, and these clusters are not included in the plot.
    \label{figure:xraysz}
  }
\end{figure*}

\begin{figure*}
  \begin{minipage}[b]{0.5\linewidth}
\includegraphics[width=6.0in]{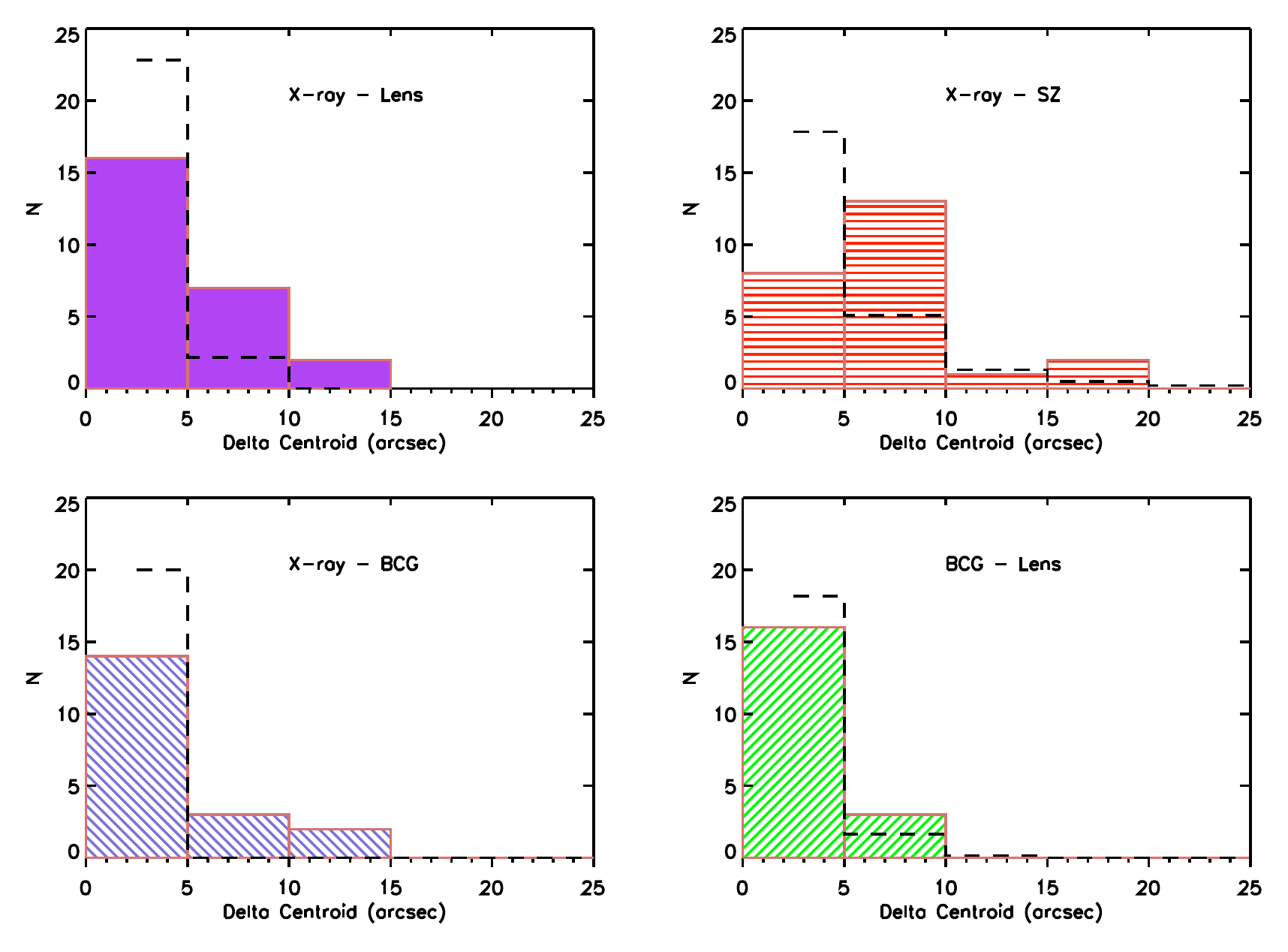}
  \end{minipage}
\caption{\footnotesize
  Histogram of centroid offsets (arcseconds). The dashed line shows the expected distribution of offsets based
  on measurement uncertainty if the true offset were identically zero.
  \label{figure:hist_cent}
  }
\end{figure*}

\begin{figure*}
  \begin{minipage}[b]{0.5\linewidth}
\includegraphics[width=6.0in]{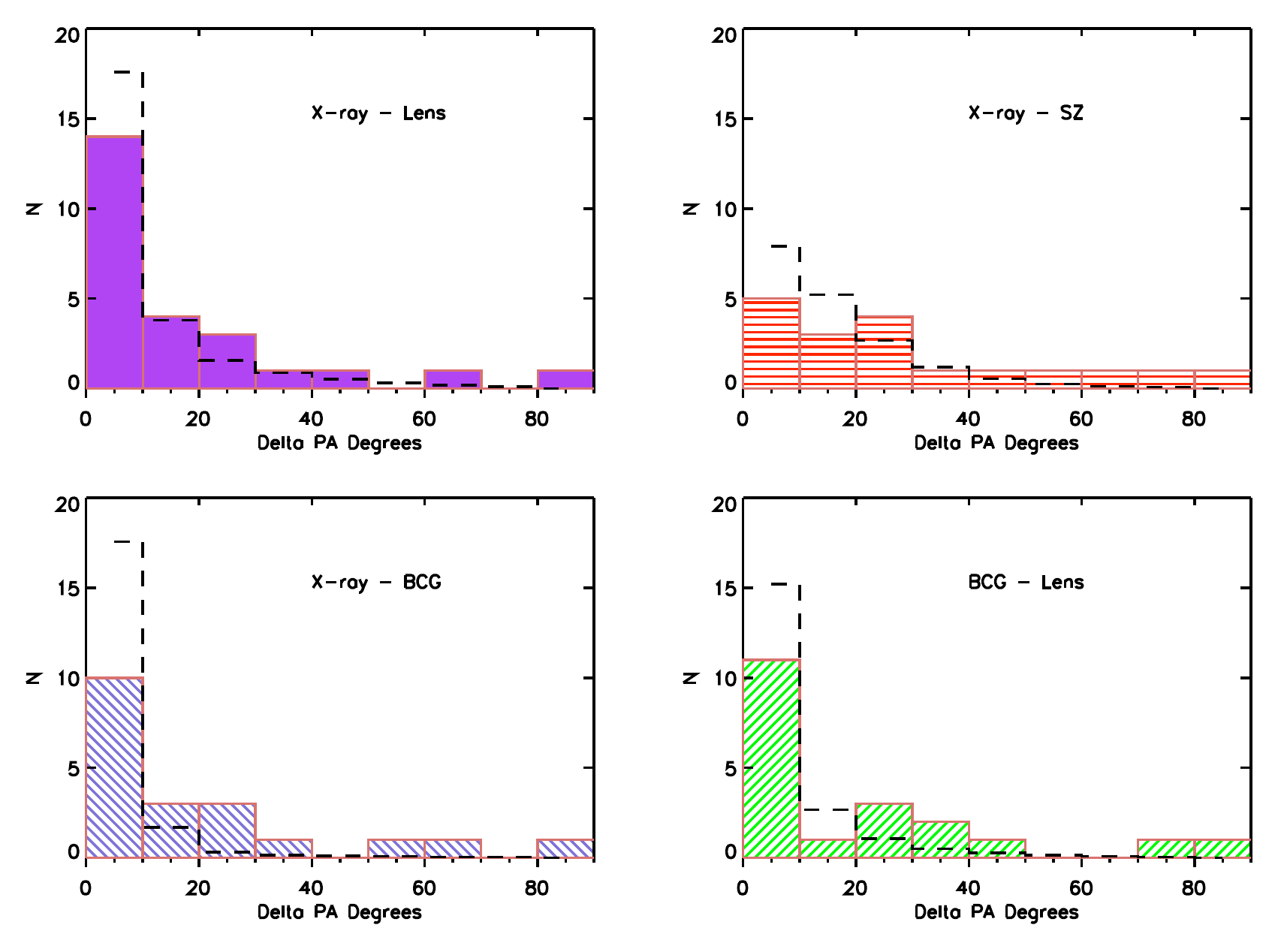}
  \end{minipage}
\caption{\footnotesize
  Histogram of PA offsets (degrees). The dashed line shows the expected distribution of offsets based
  on measurement uncertainty if the true offset were identically zero.
  \label{figure:hist_PA}
  }
\end{figure*}

\subsection{Comparison of X-ray Morphological Properties of CLASH Clusters with Simulations}

In this section we compare the X-ray morphologies of CLASH clusters to the predicted X-ray
morphology of clusters from the MUSIC simulation with similar masses and redshifts, but
no morphological selection. For convenience, we choose to show the morphological parameters measured for 
the simulated clusters at $z=0.333$. 
We verified that the results are similar for the other redshifts.
Each simulated cluster is represented three times in this sample, for 3 different viewing angles. 
Remember that in this work we are considering the complete set of simulated clusters selected only on the basis of the gravitational mass, and therefore the simulated clusters plotted in Figure~\ref{figure:MUSIC} can represent any state of relaxation and virialization. 
For this reason, the expectation for the CLASH clusters is that their morphological parameters 
will be in the range of simulated clusters but skewed. 
In particular, we expect the CLASH clusters should be rounder (i.e. axial ratios near unity), 
and to have smaller $w$ and power ratios than mean values of the simulated clusters.

Scatter plots for the parameters $C$, $w$, axial ratio, and power ratios $P30$ and $P40$
are presented in Figure~\ref{figure:MUSIC}, in which one can see that 
the distribution of simulated clusters is similar to that of 
the observed CLASH clusters in the power ratios $P30$ and $P40$. 
The two power ratios are correlated and the 5 less-relaxed, lensing-selected CLASH clusters 
have larger power ratios than the 20-object X-ray selected CLASH subsample.
On average, the CLASH clusters are rounder than the simulated clusters, 
in the sense that their axial ratios are closer to unity. 
This resultis not surprising, since the CLASH X-ray clusters were chosen in part for their round X-ray isophotes. 
However, even the irregular, lensing-selected clusters are rounder than most of the simulated clusters, 
according to the axial-ratio measurements:  All of the lensing-selected CLASH clusters have an axial ratio about 0.8, while 
only 15\% of the simulated sample has an axial ratio above 0.8.

The centroid shifts $w$ of the CLASH clusters are smaller than most
simulated clusters of similar mass, but they are not out of range: only 25\% of the
MUSIC clusters have $w \leq 0.01$ while 60\% of the CLASH sample have
such small $w$.

The range of concentration $C$ of the CLASH clusters is similar to
that of the simulated sample. Note that a significant subset of simulated clusters have
very high concentrations compared to those observed for CLASH
clusters, even for the relaxed cool core clusters in CLASH. None of the
simulated clusters have concentrations as low as a few of the CLASH
strong-lensing selected clusters.

 To summarize, compared to simulated clusters in MUSIC with a similar
mass, CLASH cluster morphologies are on average rounder and have
smaller centroid shifts. Their surface brightness concentrations and power ratios are
similar to that of the mass-selected sample of simulated clusters.

 \citet{meneghetti.etal.2014} defined regularity $M$ for a given simulated cluster in terms of the
offsets of a set of morphological properties, in units of standard deviations. 
We refer the reader to \citet{meneghetti.etal.2014} for
details and specific relations, but
we briefly review the results relevant to this work here. For each of five morphological parameters, 
$w$, eccentricity, $P30$, and $P40$, and $1/C$, they find the difference between the log of the parameter and
the log of the mean and divide by the standard deviation of the log quantities. 
They then sum these ratios to arrive at a composite regularity estimate ($M$). 
The more negative $M$ is, the smaller $w$, $P30$, and $P40$, and the larger $C$ is compared to the full sample; the
clusters with the most negative $M$ are generally rounder, more symmetric, and have higher central X-ray surface brightnesses compared to their outskirts. 
In addition, \citet{meneghetti.etal.2014}, define a simulated cluster as ``relaxed'' if the center of mass
displacement from the minimum of the gravitational potential is small ($\sim 0.07$ of the virial radius) 
and ``super-relaxed'' if in addition to a small displacement of the center of mass, 
the ratio between two times the kinetic and
gravitational energy ($2T/|U|$) is nearly unity ($<1.35$), and the mass in substructures is
small ($<10\% M_{vir}$). 
They found no correlation between X-ray regularity $M$ 
and the fraction of non-relaxed or  relaxed systems (see their Figure 15, lower panel). 
However, they saw a small correlation of $M$ with ``super-relaxed'' systems 
as one might expect: there are more ``super-relaxed'' clusters that are X-ray regular (negative $M$) 
than there are super-relaxed clusters that are X-ray irregular (positive M).
For all simulated clusters, the mean ratio $2T/|U|$  
was $1.37 \pm 0.10$ and the mean center of mass offset was $0.08 \pm 0.05 R_{vir}$
and the fractional mass in resolved substructures was $0.25 \pm 0.20$. 
For comparison, the same quantities for the simulated clusters chosen to match the CLASH clusters
(as in \citealt{meneghetti.etal.2014})   
were $1.35 \pm 0.08$, $0.06 \pm 0.04 R_{vir}$, and $0.21 \pm 0.13$, respectively. 
Thus, the relaxation measures of the complete, unabridged MUSIC runs and of the
CLASH-like sub-sample  are statistically similar, with differences in the means of a few
percent, but always consistent at 1 $\sigma$.
The X-ray morphology of both the CLASH clusters and the simulated clusters are measured 
within 500 kpc while any relaxation metric for a simulated cluster 
 extends to the virial radius. A cluster can be X-ray regular in its center 
 while having substructure in its outskirts.

\begin{figure*}
  \begin{minipage}[b]{0.5\linewidth}
\includegraphics[width=6.0in]{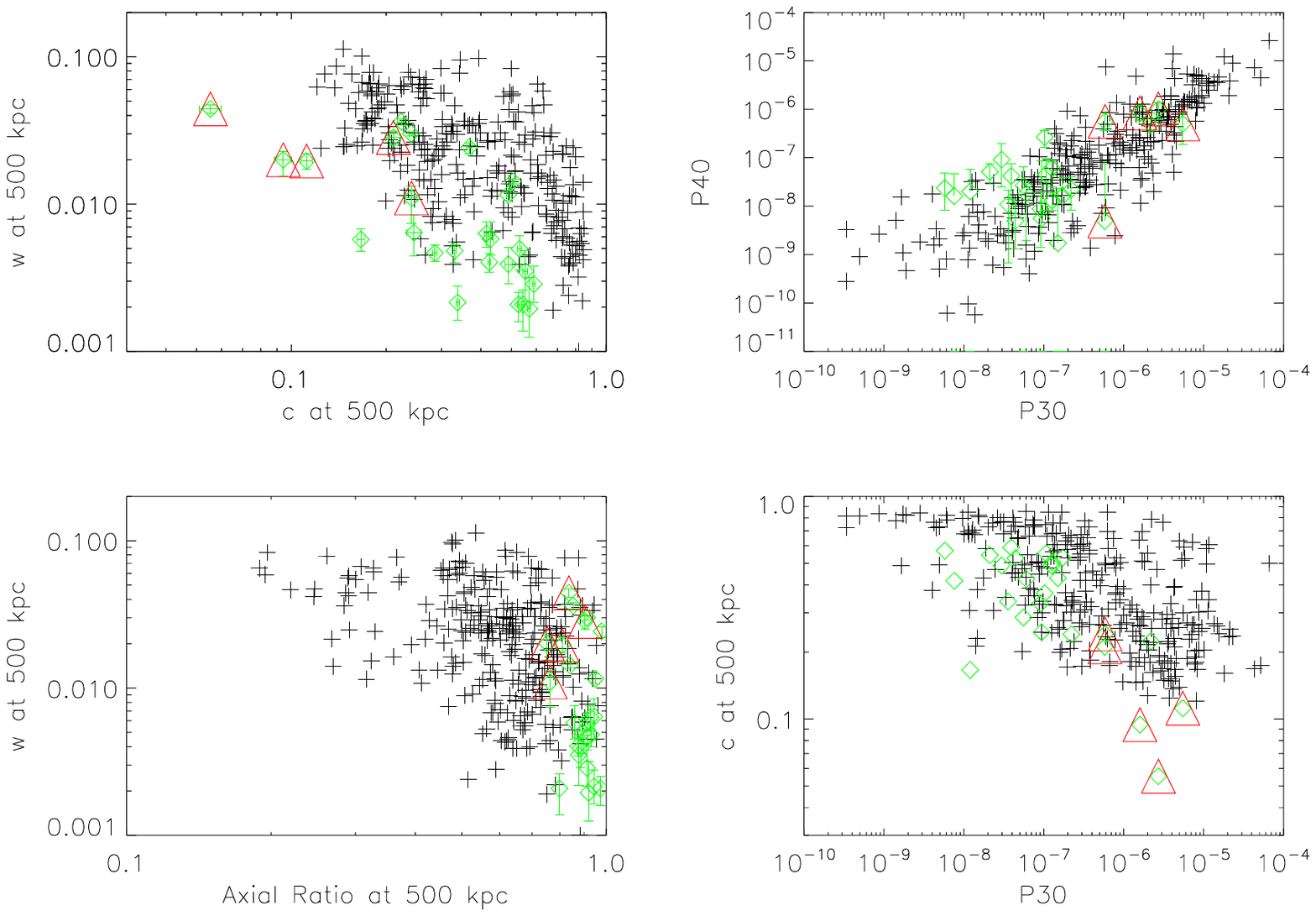}
  \end{minipage}
\caption{\footnotesize
  X-ray properties of simulated clusters from the MUSIC-2 simulation are plotted as black crosses. 
  The CLASH X-ray morphological properties are plotted with green diamonds and
  error bars. The results for the five
  lensing-selected CLASH clusters are identified on each plot with red triangles. These clusters
  are expected to be less relaxed than the 20 X-ray selected clusters of the CLASH sample.
  \label{figure:MUSIC}
  }
\end{figure*}

\section{Brightest Cluster Galaxy - Cluster Mass Alignments}

Brightest cluster galaxies (BCGs) are well-known for their alignment 
with the galaxy distribution of their host cluster of galaxies 
\citep[e.g.][]{1968PASP...80..252S, 1978ApJ...226...55D,1987ApJ...317..668S,1988AJ.....95..298T}
and with their X-ray emission \citep[e.g.][]{2008MNRAS.390.1562H,1995MNRAS.275..741A}.
This alignment indicates a connection between the galaxy-scale potential
of the BCG and the larger cluster-scale potential.
An early model of the formation of a BCG \citep{1994MNRAS.268...79W} posits that the formation of the
BCG is coupled to the formation of the cluster, and that the BCG stellar distribution retains
a memory of the preferred accretion axis for the cluster itself. This explanation is viable and
has survived observational tests such as \citet{2014MNRAS.440..588H} and \citet{2010MNRAS.405.2023N}. 
However, cosmological simulations of galaxy-cluster assembly still do not have enough spatial 
resolution to test this hypothesis, in that the spatial structure of the central galaxy on kpc scales
is not well resolved in these large-volume simulations.  

Our sample is not large enough or diverse enough to add much to what has already been discussed
about the alignment of BCGs and their host clusters as seen in optical and X-ray light. However,
the detailed lensing and SZE maps for this sample are new, and therefore we report here
a distinct BCG-cluster alignment effect between 10-kpc scale position angle measured
from the stellar distribution from HST images of the BCG, 
and the gravitational potential measured at 500-kpc by lensing, X-ray,
and SZE maps for the BCGs in the CLASH sample. This correlation is significant 
even though these systems were chosen to be relatively round in their X-ray appearance.

\citet{2015ApJ...805..177D} measured the position angle (PA) and centroid of the near-infrared, rest-frame 1-micron light in the CLASH Brightest
Cluster Galaxies (BCGs) in 
a similar fashion to the measurements presented in this work. The radial scales of the measurements, derived from the
analyses of HST WF3/NIR images were of order 10 kpc for all 25 BCGs. The apertures were chosen to avoid contamination from
other cluster galaxies and lensed features for quiescent BCGs and to overlap the regions of excess UV light in 
the others. 
The 1-micron light from BCGs is dominated by light from stars, primarily old
stars (5-10 billion years old). The gravitational mass in the centers of BCGs is also dominated by stars, so the starlight
and the mass in the BCGs might be expected to be very well aligned.

The BCGs centroids align very well with the X-ray centroids of CLASH clusters, which is not surprising because 
good X-ray alignment with the BCG was a selection criterion for the 20 X-ray selected clusters. 
The typical PA difference between the BCG and the X-ray PAs is $2 \pm 24$ degrees, while
between BCG and lensing PAs is $5 \pm 25$ degrees. The  PA difference between BCG and the cluster SZE PA is $38 \pm 22$ degrees. 
The standard Spearman's test indicates a strong correlation in all 3 of these comparisons, where low
probability indicates high significance of correlation: 0.89 (probability $=2 \times 10^{-7}$) for X-ray/BCG,
0.81 (probability $=1.5 \times 10^{-7}$) for lensing/BCG, and 0.83 (probability=$2 \times 10^{-5}$) for SZE/BCG position angles. 
The 11 clusters with low central entropies ($K_0 < 30$ keV cm$^{2}$), or
cool core clusters, show less dispersion: $4 \pm 10$, $4 \pm 14$ , and $33 \pm 18$ degrees respectively, for the X-ray, lensing and 
SZE- determined position 
angles. The correlation is somewhat less significant in the cool core sample because of the smaller number of clusters, 
but similarly strong ($r=0.79-0.75$ with probability $=0.004-0.007$ for x-ray-BCG and lens-BCG
alignments respectively, while the SZ-BCG correlation drops to $r=0.60$ with probability $=0.07$, indicating a less
than $2-\sigma$ correlation for PA in the CC-SZE BCG sample.

The offsets are correlated between lensing and X-ray comparisons, in that the
BCGs with the largest X-ray PA offsets have the largest lensing PA offsets as well (Figure~\ref{figure:BCG}).
The largest outliers in the X-ray/BCG comparison are Abell 2261 and MACS1206. Both of these clusters are BCG-dominated,
non-cool core systems. Abell 2261 is also a major outlier in the lensing/BCG PA comparison, while the BCG in MACS1206
is well-aligned with the lensing map. MACS0744 is the other significant outlier in the lensing/BCG PA comparison,
(in X-rays, the PA for MACS0744 is not well determined.) MACS0744 is also a BCG-dominated, non-cool core cluster.

 In summary, the PAs of the near-infrared light of BCGs aligns very well with the PAs of the X-ray gas maps, SZE maps, and the lensing projected mass maps. That alignment is not trivial, because of the factor of 50 difference in the radii where the 
PA is defined and compared for the BCG with that of the larger-scales of the cluster.
The PA of the BCG is measured at a scale of 10 kpc or less, and the PA of the gas
and the projected mass were measured at a scale of 500 kpc. The correlation suggests that 
the mass distribution at 500 kpc is strongly coupled to the mass distribution
at 10 kpc, even in these relatively round and relaxed systems.

\begin{figure*}
  \begin{minipage}[b]{0.5\linewidth}
\includegraphics[width=6.0in]{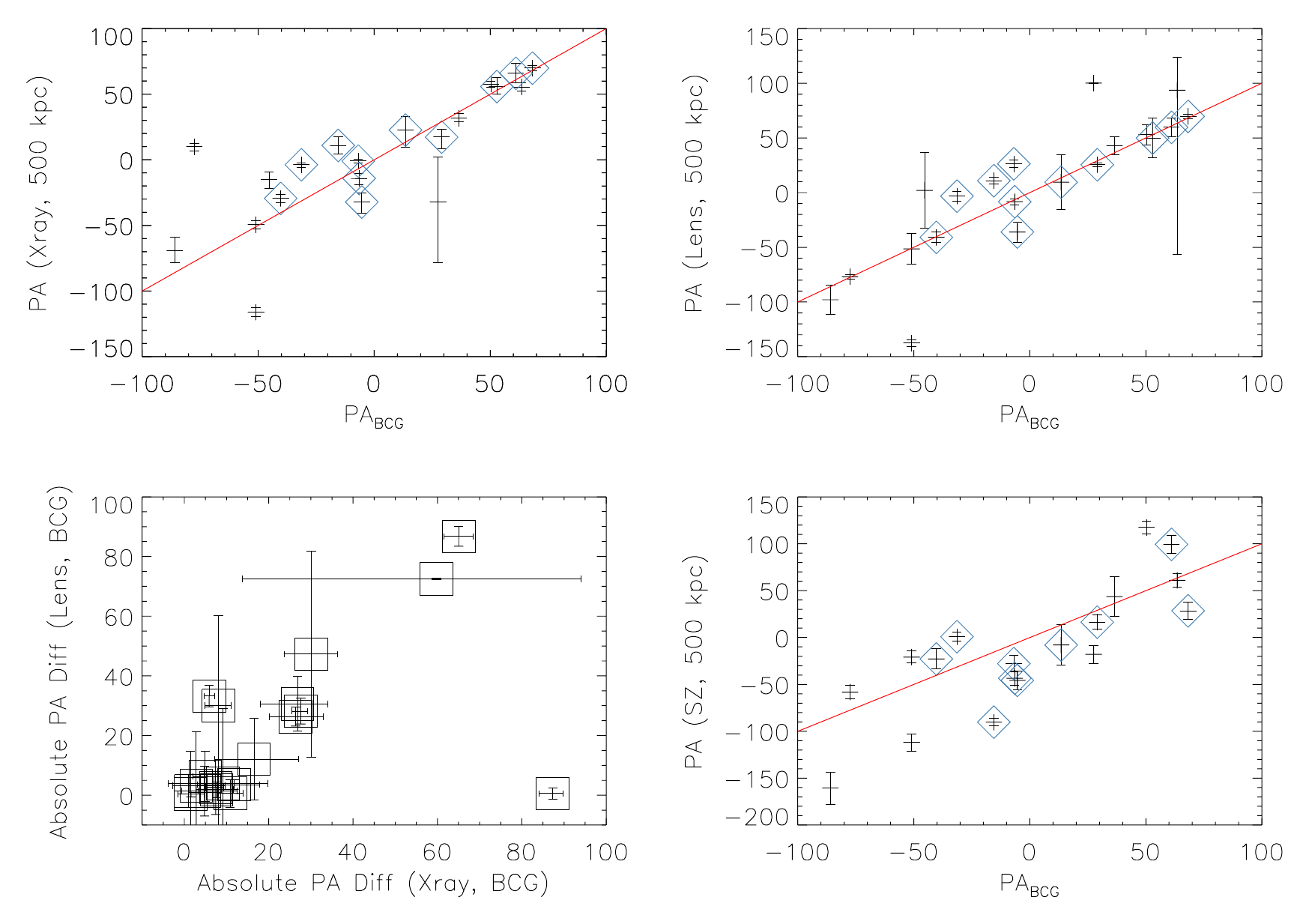}
  \end{minipage}
\caption{\footnotesize
  (Top Row) Position Angle in degrees of the near-IR emission of the BCG from \citet{2014ApJ...794..136D} plotted against 
  the PA for the X-ray and lensing maps at 500 kpc. (Lower Left) The absolute value of the PA difference for the 
  BCG, X-ray reference on the x-axis, while the lens PA reference is on the y-axis. The dispersion from zero offset is
  about 25 degrees. The most significant outlier here is MACS1206, which has almost no offset between the PA measured
  in lensing and for the BCG, but the X-ray PA is almost perpendicular to those two. (Lower Right) The SZE position angle in degrees plotted against the PA of the BCG.
  \label{figure:BCG}
  }
\end{figure*}

\section{Conclusions}

The CLASH project \citep{2012ApJ...756..159P} has collected a significant amount of data for a sample of 25 massive clusters of galaxies
with redshifts from 0.2-0.9, including strong and weak lensing
constraints from HST \citep{2015ApJ...806....4M}, weak lensing constraints from Subaru \citep{2014ApJ...795..163U}, 
X-ray observations from Chandra and XMM \citep{2014ApJ...794..136D}, and SZE observations from Bolocam \citep{Czakon2015}.
To compare the results of the CLASH cluster sample 
to predictions from simulations, \citet{meneghetti.etal.2014} selected simulated clusters
replicating the morphological and temperature selection of the original CLASH sample. 
We present here the full uniformly measured X-ray morphological parameters and uncertainties for the 25 CLASH clusters used
in that work. In addition we measure the same parameters in lensing and SZE maps. All parameters are measured inside radii of 500 kpc, and
for a sub-sample of the clusters inside $0.5 R_{500}$. The full set of morphological 
properties are centroids, concentration ratios ($C$), centroid shifts ($w$), axial ratios, power
ratios ($P_3/P_0$, $P_4/P_0$), and position angles.
We present means and standard deviations of these  properties for the CLASH sample and for the X-ray selected CLASH subsample.
For the first time we demonstrate a strong correlation between these morphological quantities as measured from 
lensing, X-ray, and SZE maps at a consistent radial scale of 500 $h_{70}^{-1}$ kpc (which is about half of $R_{500}$ for these clusters of galaxies.)

In order to visualize how typical CLASH clusters are relative to a complete set of simulated clusters
of similar mass, 
we compared high-mass clusters from the MUSIC simulations \citep{2014ApJ...797...34M}
with our CLASH observations. The simulated clusters were mass-selected but are not matched in morphology 
as they were in \citet{2014ApJ...797...34M}.
The full set of simulated clusters were expected to be more heterogeneous in structure and dynamic state in terms of relaxation than the CLASH sample. 
The CLASH clusters have similar power ratios, but $C$, $w$, and AR indicate
that the CLASH clusters indeed appear more regular than a typical simulated cluster of similar mass. 
Very early attempts to simulate X-ray cluster morphologies typically failed to create 
clusters that were as relaxed as those selected from X-ray observations. 
The first simulations had limited cosmological context (simulations of individual 
clusters) or incorrect cosmology, such as 
a standard CDM universe with $\Omega_M=1$ which predicts more recent assembly activity, 
as discussed in \citep{1996MNRAS.282...77T,1997MNRAS.284..439B}.  Some early simulations assumed 
lower matter densities  \citep{1993ApJ...419L...9E,1995ApJ...447....8M} and late time acceleration of the expansion \citep{1998MNRAS.296.1061T}, 
but accurate predictions 
of the distribution of hot gas inside of clusters require baryon prescriptions including
the effects of AGN and star formation feedback \citep{rasia.etal.2015}.
These processes regulate the cooling and the heating of the cluster core and have significant impact
on the central region. The morphological parameter most affected by baryon processes is the X-ray surface brightness concentration.
However, AGN feedback may also change the shape and distribution of substructures because gas stripping 
becomes more efficient and merging systems are thermalized more quickly, potentially leading to changes in axis ratios.  
Future work, expanding beyond what the CLASH team reported in \citet{2014ApJ...797...34M}, will be able to compare simulations of high-mass clusters of galaxies to 
these well-characterized CLASH clusters can be made by selecting 
clusters based on estimated morphological
properties and mass or gas temperatures similar to those of CLASH clusters.

We also show that the stellar mass of the BCG at small scales (10 kpc), is strongly aligned with the position 
angle of the matter distribution on much larger scales (500 kpc), probed by lensing, X-ray, and SZE observations.  
To our knowledge, this the first time BCG position angle has been demonstrated to align with the position angle
of the mass distribution in a galaxy-cluster sample with detailed lensing maps.  However, the result has its roots in mid-20th century astronomy.  Alignments of BCGs with the galaxy distribution of their host clusters have been noted by Abell and others since the 1950s. 

This correlation shows that there is a strong relationship between the assembly of the 
mass distribution of the stars in the very center of the matter halo, inside the brightest cluster galaxy, 
and the distribution of dark matter and the hot X-ray at large scales.
The underlying cause of this correlation is likely to be the shape of the underlying gravitational potential. 
Cluster potential shapes are not expected to be perfectly spherical because
in models of large scale structure matter does not flow in evenly over 4$\pi$ steradians but preferentially along filaments. 
Systematic alignment of BCGs with their clusters suggests that even the very inner regions of a galaxy cluster 
reflect the anisotropy of mass accretion on much larger scales.

\begin{deluxetable*}{lllllllllllllll}
\tablefontsize{\footnotesize}
\tablecaption{X-ray Morphological Properties for the CLASH Sample  500 kpc \label{table:xray500morph}}
\tablehead{\colhead{Name}         & \colhead{RA} & \colhead{Dec} & \colhead{$C$ } & \colhead{$C$ } & \colhead{$w$} & \colhead{$w$ }   & \colhead{P30}     & \colhead{P30 } & \colhead{P40}     & \colhead{P40 } & \colhead{AR} & \colhead{AR}  & \colhead{PA}  & \colhead{PA} \\
\colhead{}         & \colhead{centroid} & \colhead{centroid} & \colhead{} & \colhead{unc} & \colhead{} & \colhead{unc}   & \colhead{}     & \colhead{unc} & \colhead{}     & \colhead{unc} & \colhead{}  & \colhead{unc} & \colhead{}  & \colhead{unc}}
\startdata
A209      & 22.97016  & -13.6118  & 0.167 & 0.005 & 5.2E-03 & 1.0E-03 & 4.8E-08 & 5.6E-08 & 3.78E-08 & 3.3E-08 & 0.91 & 0.01 & -49 & 3  \\
A383      & 42.01374  & -3.52945  & 0.525 & 0.004 & 1.8E-03 & 4.7E-04 & 5.8E-08 & 2.7E-08 & 6.03E-09 & 6.1E-09 & 0.97 & 0.01 & 22  & 11 \\
0329-02   & 52.42264  & -2.19577  & 0.488 & 0.006 & 1.1E-02 & 1.1E-03 & 1.6E-07 & 8.5E-08 & 6.76E-08 & 3.8E-08 & 0.95 & 0.02 & -31 & 6  \\
0416-24   & 64.03614  & -24.07162 & 0.091 & 0.004 & 1.5E-02 & 4.0E-03 & 1.5E-06 & 7.4E-07 & 8.54E-07 & 3.5E-07 & 0.75 & 0.02 & 37  & 2  \\
0429-02   & 67.39994  & -2.88486  & 0.531 & 0.006 & 4.8E-03 & 1.4E-03 & 2.0E-07 & 1.0E-07 & 2.67E-08 & 2.0E-08 & 0.89 & 0.02 & -14 & 4  \\
0647+70   & 101.95787 & 70.24891  & 0.242 & 0.008 & 1.0E-02 & 2.3E-03 & 6.4E-07 & 3.1E-07 & 5.68E-07 & 2.8E-07 & 0.76 & 0.02 & -80 & 2  \\
0717+37   & 109.38513 & 37.75319  & 0.055 & 0.004 & 4.5E-02 & 3.4E-03 & 2.5E-06 & 9.7E-07 & 9.75E-07 & 4.1E-07 & 0.84 & 0.02 & -29 & 2  \\
0744+39   & 116.21812 & 39.45748  & 0.365 & 0.006 & 2.4E-02 & 1.6E-03 & 1.4E-07 & 1.2E-07 & 2.81E-07 & 1.0E-07 & 0.98 & 0.01 & -22 & 37 \\
A611      & 120.23689 & 36.05725  & 0.335 & 0.004 & 2.0E-03 & 4.6E-04 & 8.0E-08 & 4.0E-08 & 2.33E-08 & 1.6E-08 & 0.95 & 0.01 & 33  & 4  \\
1115+01   & 168.96666 & 1.49813   & 0.430 & 0.008 & 5.6E-03 & 1.7E-03 & 1.9E-07 & 1.2E-07 & 1.68E-08 & 2.5E-08 & 0.85 & 0.03 & -29 & 3  \\
1149+22   & 177.39769 & 22.4004   & 0.111 & 0.007 & 1.9E-02 & 3.2E-03 & 5.5E-06 & 1.6E-06 & 5.27E-07 & 3.3E-07 & 0.81 & 0.02 & -36 & 2  \\
A1423     & 179.32236 & 33.61042  & 0.287 & 0.004 & 4.6E-03 & 7.0E-04 & 6.8E-08 & 5.4E-08 & 4.13E-08 & 3.0E-08 & 0.91 & 0.02 & 55  & 3 \\
1206-08   & 181.55339 & -8.8027   & 0.223 & 0.007 & 3.6E-02 & 2.0E-03 & 2.3E-06 & 4.4E-07 & 5.29E-07 & 1.4E-07 & 0.85 & 0.01 & 10  & 2  \\
1226+33   & 186.74102 & 33.54674  & 0.245 & 0.010 & 5.5E-03 & 2.1E-03 & 1.9E-07 & 2.5E-07 & 6.05E-08 & 7.4E-08 & 0.95 & 0.03 & -69 & 13 \\
1311-03   & 197.75704 & -3.17733  & 0.488 & 0.013 & 3.2E-03 & 1.2E-03 & 1.0E-07 & 1.1E-07 & 1.10E-07 & 7.4E-08 & 0.89 & 0.04 & -16 & 6  \\
1347-1145 & 206.87852 & -11.753   & 0.506 & 0.003 & 1.4E-02 & 3.6E-04 & 1.3E-07 & 2.5E-08 & 1.48E-08 & 7.2E-09 & 0.84 & 0.01 & -1  & 1  \\
1423+24   & 215.94964 & 24.07839  & 0.555 & 0.009 & 3.5E-03 & 1.4E-03 & 1.6E-07 & 1.2E-07 & 6.20E-08 & 4.9E-08 & 0.88 & 0.02 & 17  & 7  \\
1532+30   & 233.22438 & 30.34978  & 0.571 & 0.007 & 1.9E-03 & 8.3E-04 & 1.2E-08 & 1.3E-08 & 2.64E-08 & 1.8E-08 & 0.92 & 0.02 & 55  & 6  \\
1720+35   & 260.06941 & 35.60649  & 0.417 & 0.008 & 6.1E-03 & 1.3E-03 & 2.8E-08 & 3.7E-08 & 2.52E-08 & 2.3E-08 & 0.93 & 0.02 & 11  & 6  \\
A2261     & 260.61273 & 32.13266  & 0.331 & 0.004 & 4.7E-03 & 5.7E-04 & 1.2E-07 & 4.2E-08 & 1.25E-08 & 9.4E-09 & 0.93 & 0.01 & 63  & 3  \\
1931-26   & 292.95663 & -26.57594 & 0.545 & 0.006 & 1.7E-03 & 6.2E-04 & 3.5E-08 & 2.8E-08 & 6.56E-08 & 2.3E-08 & 0.80 & 0.01 & -4  & 2  \\
2129-07   & 322.3573  & -7.69189  & 0.211 & 0.010 & 2.7E-02 & 2.8E-03 & 7.0E-07 & 4.0E-07 & 4.85E-08 & 5.6E-08 & 0.90 & 0.02 & 84  & 6  \\
2129+0005 & 322.41548 & 0.08858   & 0.426 & 0.004 & 3.8E-03 & 5.7E-04 & 6.1E-08 & 2.8E-08 & 2.45E-08 & 1.2E-08 & 0.87 & 0.01 & 70  & 2  \\
MS2137    & 325.0635  & -23.66098 & 0.589 & 0.007 & 2.5E-03 & 7.9E-04 & 5.1E-08 & 5.7E-08 & 5.60E-08 & 3.7E-08 & 0.92 & 0.03 & 66  & 8  \\
2248-44   & 342.18683 & -44.52922 & 0.194 & 0.003 & 1.5E-02 & 9.7E-04 & 2.0E-07 & 7.1E-08 & 2.74E-08 & 1.6E-08 & 0.91 & 0.01 & 62  & 2  \\
\enddata
\tablecomments{ The uncertainty range reported in this table and Table~\ref{table:xray05r500morph} is computed from the difference between the upper and lower 68 percent ($\sim 1 \sigma$) of the fits after 100 Monte Carlo runs, and the reported morphology value is the median value from those runs. AR is axis ratio; PA is Position Angle, in degrees east of north.}
\end{deluxetable*}

\begin{deluxetable*}{lllllllllllllll}
\tablefontsize{\footnotesize}
\tablecaption{X-ray Morphological Properties for the CLASH Sample $0.5 R_{500}$ \label{table:xray05r500morph}}
\tablehead{\colhead{Name}         & \colhead{RA} & \colhead{Dec} & \colhead{$C$ } & \colhead{$C$ } & \colhead{$w$} & \colhead{$w$ }   & \colhead{P30}     & \colhead{P30 } & \colhead{P40}     & \colhead{P40 } & \colhead{AR} & \colhead{AR}  & \colhead{PA}  & \colhead{PA} \\
\colhead{}         & \colhead{centroid} & \colhead{centroid} & \colhead{} & \colhead{unc} & \colhead{} & \colhead{unc}   & \colhead{}     & \colhead{unc} & \colhead{}     & \colhead{unc} & \colhead{}    & \colhead{unc} & \colhead{} & \colhead{unc}}
\startdata
A209     & 22.96919     & -13.6111      & 0.215     & 0.005  & 5.0E-03            & 1.0E-03 & 2.8E-08  & 3.0E-08 & 1.9E-08  & 1.8E-08 & 0.88    & 0.02 & -46.2  & 2.5  \\
A383     & 42.01382     & -3.52943      & 0.594     & 0.004  & 1.6E-03            & 4.1E-04 & 5.2E-09  & 5.5E-09 & 1.4E-08  & 8.2E-09 & 0.98    & 0.01 & -11.5  & 19.4 \\
0329-02 & 52.42257     & -2.1957       & 0.511     & 0.006  & 1.2E-02            &  1.1E-03 & 1.5E-07  & 6.5E-08 & 2.4E-08  & 1.8E-08 & 0.91    & 0.02 & -28.3  & 5.0  \\
0416-24 & \nodata      & \nodata      & \nodata    & \nodata  & \nodata           & \nodata & \nodata & \nodata & \nodata & \nodata & \nodata & \nodata & \nodata & \nodata \\
0429-02 & 67.40009     & -2.88479      & 0.563     & \nodata  & 4.7E-03         &   1.1E-03 & 4.4E-08  & 4.6E-08 & 9.3E-09  & 1.3E-08 & 0.89    & 0.02 & -11.9  & 4.7  \\
0647+70 & \nodata      & \nodata       & \nodata   & \nodata  & \nodata           & \nodata & \nodata & \nodata & \nodata & \nodata & \nodata & \nodata & \nodata & \nodata \\
0717+37 & \nodata      & \nodata       & \nodata   & \nodata  & \nodata           & \nodata & \nodata & \nodata & \nodata & \nodata & \nodata & \nodata & \nodata & \nodata \\
0744+39 & 116.21804    & 39.45748      & 0.375     & 0.007  & 2.3E-02            & 1.6E-03 & 1.5E-07  & 1.1E-07 & 2.0E-07  & 8.2E-08 & 0.98    & 0.01 & -18.9  & 30.7 \\
A611     & 120.23706    & 36.05736      & 0.394     & 0.004  & 2.4E-03            & 7.3E-04 & 7.9E-09  & 9.4E-09 & 4.6E-09  & 4.8E-09 & 0.94    & 0.01 & 40.7   & 5.1  \\
1115+01 & 168.96671    & 1.49809       & 0.481     & 0.008  & 5.3E-03            & 1.8E-03 & 8.8E-08  & 7.7E-08 & 1.0E-08  & 1.2E-08 & 0.87    & 0.02 & -27.4  & 5.1  \\
1149+22 & \nodata      & \nodata       & \nodata   & \nodata  & \nodata           & \nodata & \nodata & \nodata & \nodata & \nodata & \nodata & \nodata & \nodata & \nodata \\
A1423    & \nodata      & \nodata       & \nodata   & \nodata  & \nodata           & \nodata & \nodata & \nodata & \nodata & \nodata & \nodata & \nodata & \nodata & \nodata \\
1206-08 & 181.55361    & -8.80286      & 0.262     & 0.006  & 3.2E-02            & 1.4E-03 & 1.5E-06  & 3.2E-07 & 2.0E-07  & 7.2E-08 & 0.85    & 0.01 & 9.0    & 2.2  \\
1226+33 & 186.74112    & 33.54693      & 0.315     & 0.011  & 6.8E-03            & 1.6E-03 & 4.0E-07  & 3.0E-07 & 6.9E-08  & 5.9E-08 & 0.93    & 0.02 & -86.3  & 10.5 \\
1311-03 & 197.75701    & -3.17731      & 0.477     & 0.012  & 3.9E-03            & 1.1E-03 & 7.9E-08  & 1.0E-07 & 1.0E-07  & 9.1E-08 & 0.86    & 0.03 & -14.4  & 5.8  \\
1347-1145 & 206.87867    & -11.75312     & 0.585     & 0.003  & 1.2E-02            & 3.6E-04 & 5.1E-08  & 1.4E-08 & 1.4E-08  & 4.8E-09 & 0.85    & 0.01 & 0.4    & 1.2  \\
1423+24 & 215.94964    & 24.07835      & 0.562     & 0.009  & 4.7E-03            & 1.4E-03 & 6.9E-08  & 7.1E-08 & 3.1E-08  & 3.3E-08 & 0.89    & 0.03 & 18.0   & 7.0  \\
1532+30 & 233.22436    & 30.34977      & 0.587     & 0.007  & 2.0E-03            & 7.9E-04 & 2.0E-08  & 2.0-E-08 & 7.7E-09  & 8.4E-09 & 0.93    & 0.02 & 51.7   & 9.0  \\
1720+35 & 260.06927    & 35.60642      & 0.458     & 0.007  & 5.5E-03            & 1.2E-03 & 5.4E-08  & 4.9E-08 & 2.0E-08  & 2.0E-08 & 0.94    & 0.02 & 22.4   & 5.2  \\
A2261    & 260.6115     & 32.13223      & 0.431     & 0.004  & 5.2E-03            & 5.0E-04 & 5.7E-07  & 8.5E-08 & 2.1E-07  & 3.4E-08 & 0.86    & 0.01 & 73.9   & 1.4  \\
1931-26 & 292.95663    & -26.57591     & 0.575     & 0.005  & 1.8E-03            & 5.9E-04 & 4.0E-08  & 2.7E-08 & 2.4E-08  & 1.6E-08 & 0.79    & 0.01 & -4.2   & 1.8  \\
2129-07 & \nodata      & \nodata       & \nodata   & \nodata  & \nodata           & \nodata & \nodata & \nodata & \nodata & \nodata & \nodata & \nodata & \nodata & \nodata \\
2129+00 & 322.41523    & 0.08861       & 0.461     & 0.004  & 4.8E-03            & 8.1E-04 & 3.9E-08  & 2.2E-08 & 3.2E-08  & 1.4E-08 & 0.87    & 0.01 & 68.8   & 1.9  \\
MS2137  & 325.06373    & -23.66072     & 0.612     & 0.007  & 5.4E-03            & 9.9E-04 & 3.1E-08  & 3.4E-08 & 2.5E-08  & 2.2E-08 & 0.97    & 0.03 & 60.3   & 7.3  \\
2248-44 & 342.18634    & -44.52951     & 0.333     & 0.003  & 2.2E-02            & 7.1E-04 & 1.8E-08  & 1.3E-08 & 5.2E-09  & 4.7E-09 & 0.91    & 0.01 & 55.2   & 2.3 
\enddata
\tablecomments{ Same convention as previous Table.  }
\end{deluxetable*}

\begin{deluxetable*}{lllllllllllllll}
\tablefontsize{\footnotesize}
\tablecaption{Lensing Morphological Properties for the CLASH Sample 500 kpc \label{table:lens500morph}}
\tablehead{\colhead{Name}         & \colhead{RA} & \colhead{Dec} & \colhead{$C$ } & \colhead{$C$ } & \colhead{$w$} & \colhead{$w$ }   & \colhead{P30}     & \colhead{P30 } & \colhead{P40}     & \colhead{P40 } & \colhead{AR} & \colhead{AR}  & \colhead{PA}  & \colhead{PA} \\
\colhead{}         & \colhead{centroid} & \colhead{centroid} & \colhead{} & \colhead{unc} & \colhead{} & \colhead{unc}   & \colhead{}     & \colhead{unc} & \colhead{}     & \colhead{unc} & \colhead{} & \colhead{unc}  & \colhead{}  & \colhead{unc} }
\startdata
A209      & 22.9688   & -13.6123  & 0.18 & 0.04 & 8.40E-03 & 9E-03 & 6E-08 & 4E-08 & 6E-08 & 5E-07 & 0.83 & 0.16 & -49 & 14 \\
A383      & 42.01387  & -3.52979  & 0.31 & 0.09 & 2.96E-03 & 6E-03 & 6E-08 & 2E-08 & 6E-08 & 1E-07 & 0.91 & 0.08 & 13  & 25 \\
0329-02   & 52.42153  & -2.19543  & 0.14 & 0.01 & 2.80E-02 & 1E-02 & 7E-07 & 6E-08 & 7E-07 & 3E-08 & 0.84 & 0.07 & -36 & 9  \\
0416-24   & 64.03542  & -24.07298 & 0.13 & 0.03 & 2.68E-03 & 3E-03 & 5E-07 & 3E-07 & 5E-07 & 3E-07 & 0.78 & 0.13 & 42  & 4  \\
0429-02   & 67.40018  & -2.8851   & 0.20 & 0.08 & 1.13E-03 & 2E-03 & 2E-09 & 1E-08 & 2E-09 & 8E-08 & 0.78 & 0.15 & -8  & 3  \\
0647+70   & 101.95882 & 70.24866  & 0.14 & 0.03 & 1.99E-03 & 4E-03 & 3E-08 & 2E-07 & 3E-08 & 5E-07 & 0.73 & 0.18 & -78 & 2  \\
0717+37   & 109.386   & 37.7517   & 0.04 & 0.01 & 1.56E-03 & 4E-03 & 2E-08 & 3E-07 & 2E-08 & 2E-07 & 0.94 & 0.10 & 89  & 39 \\
0744+39   & 116.21744 & 39.45835  & 0.19 & 0.06 & 3.90E-02 & 2E-02 & 3E-06 & 2E-06 & 3E-06 & 7E-07 & 0.73 & 0.19 & -80 & 0  \\
A611      & 120.23688 & 36.05659  & 0.19 & 0.07 & 1.62E-03 & 9E-03 & 4E-09 & 8E-08 & 4E-09 & 5E-08 & 0.85 & 0.14 & 43  & 8  \\
1115+01   & 168.96545 & 1.49939   & 0.15 & 0.04 & 9.32E-03 & 5E-03 & 2E-08 & 2E-09 & 2E-08 & 3E-09 & 0.87 & 0.07 & -41 & 5  \\
1149+22   & 177.399   & 22.3991   & 0.08 & 0.00 & 9.52E-03 & 2E-04 & 1E-07 & 5E-08 & 1E-07 & 2E-08 & 0.95 & 0.00 & -34 & 0  \\
A1423     & 179.32243 & 33.61049  & 0.31 & 0.09 & 1.43E-03 & 5E-03 & 1E-08 & 4E-09 & 1E-08 & 7E-08 & 0.71 & 0.28 & -88 & 30 \\
1206-08   & 181.55045 & -8.80092  & 0.13 & 0.00 & 3.01E-03 & 8E-03 & 3E-08 & 2E-07 & 3E-08 & 1E-07 & 0.80 & 0.09 & -77 & 2  \\
1226+33   & 186.74152 & 33.54608  & 0.13 & 0.01 & 3.01E-02 & 1E-02 & 3E-07 & 1E-07 & 3E-07 & 6E-08 & 0.77 & 0.09 & 82  & 14 \\
1311-03   & 197.758   & -3.17763  & 0.15 & 0.04 & 1.36E-03 & 4E-03 & 1E-09 & 2E-09 & 1E-09 & 1E-08 & 0.86 & 0.14 & 1   & 35 \\
1347-1145 & 206.879   & -11.753   & 0.14 & 0.04 & 1.41E-02 & 9E-03 & 9E-08 & 6E-08 & 9E-08 & 6E-08 & 0.81 & 0.10 & 26  & 4  \\
1423+24   & 215.94957 & 24.07864  & 0.17 & 0.06 & 1.91E-03 & 9E-03 & 7E-09 & 8E-09 & 7E-09 & 1E-07 & 0.79 & 0.14 & 26  & 2  \\
1532+30   & 233.22498 & 30.35002  & 0.19 & 0.06 & 8.91E-03 & 1E-02 & 5E-08 & 2E-08 & 5E-08 & 1E-07 & 0.84 & 0.11 & 39  & 18 \\
1720+35   & 260.06976 & 35.60713  & 0.25 & 0.11 & 2.06E-03 & 1E-02 & 8E-09 & 2E-09 & 8E-09 & 3E-07 & 0.74 & 0.21 & 11  & 3  \\
A2261     & 260.61337 & 32.13261  & 0.22 & 0.03 & 1.06E-03 & 3E-04 & 1E-08 & 1E-08 & 1E-08 & 1E-08 & 0.89 & 0.08 & 42  & 3  \\
1931-26   & 292.957   & -26.5758  & 0.18 & 0.03 & 5.56E-04 & 7E-04 & 2E-09 & 4E-09 & 2E-09 & 5E-07 & 0.70 & 0.18 & -3  & 4  \\
2129-07   & 322.35901 & -7.69128  & 0.15 & 0.04 & 2.05E-03 & 4E-03 & 2E-08 & 9E-09 & 2E-08 & 4E-07 & 0.74 & 0.15 & 86  & 3  \\
2129+0005 & 322.41678 & 0.08953   & 0.21 & 0.05 & 3.14E-03 & 9E-03 & 5E-10 & 7E-10 & 5E-10 & 6E-07 & 0.68 & 0.23 & 68  & 2  \\
MS2137    & 325.06329 & -23.65998 & 0.23 & 0.03 & 1.27E-02 & 1E-03 & 2E-08 & 2E-08 & 2E-08 & 7E-09 & 0.88 & 0.08 & 59  & 8  \\
2248-44   & 342.18338 & -44.53075 & 0.15 & 0.04 & 9.49E-04 & 1E-02 & 5E-08 & 4E-07 & 5E-08 & 8E-07 & 0.71 & 0.17 & 53  & 9 
\enddata
\tablecomments{ The morphological parameter values are based on the median. 
The uncertainty range reported in this table  is systematic because that uncertainty dwarfs the formal statistical
uncertainty in this analysis of the lensing maps. It is computed from the difference between best fit values 
based on the two lensing model assumptions
discussed in the text. The units and parameters in this table are the same as for the previous two tables.}
\end{deluxetable*}

\begin{deluxetable*}{lrrrrrrrrrr}
\tablefontsize{\footnotesize}
\tablecaption{SZE Morphological Properties for the CLASH Sample 500 kpc \label{table:SZ500morph}}
\tablehead{\colhead{Name} & \colhead{RA} & \colhead{Dec} & \colhead{$C$} & \colhead{$C$} & \colhead{$w$} & \colhead{$w$} & \colhead{AR} & \colhead{AR} & \colhead{PA} & \colhead{PA} \\
\colhead{} & \colhead{centroid} & \colhead{centroid} & \colhead{} & \colhead{unc} & \colhead{} & \colhead{unc} & \colhead{} & \colhead{unc} & \colhead{} & \colhead{unc}} \\
\startdata
A209       &  22.9705 & -13.6121 & 0.087 & 0.009 &  2.9E-03 & 3.9E-03 & 0.93 & 0.03 & -20.8 & 12.6 \\
A383       &  42.0142 &  -3.5302 & 0.057 & 0.005 & -2.0E-04 & 3.4E-03 & 0.98 & 0.01 &  -8.0 & 43.4 \\
0329-02    &  52.4222 &  -2.1972 & 0.074 & 0.012 &  1.5E-02 & 6.1E-03 & 0.89 & 0.04 & -45.7 & 20.0 \\
0416-24    &  64.0369 & -24.0708 & 0.054 & 0.013 &  9.2E-03 & 7.4E-03 & 0.93 & 0.05 &  13.1 & 15.6 \\
0429-02    &  67.3995 &  -2.8854 & 0.049 & 0.009 &  3.4E-03 & 4.5E-03 & 0.95 & 0.04 & -43.5 & 15.4 \\
0647+70    & 101.9593 &  70.2492 & 0.050 & 0.011 &  2.7E-03 & 5.5E-03 & 0.93 & 0.04 &  \nodata & \nodata \\
0717+37    & 109.3847 &  37.7518 & 0.077 & 0.012 &  1.8E-02 & 4.8E-03 & 0.95 & 0.03 &  18.8 & 41.0 \\
0744-39    & 116.2200 &  39.4582 & 0.116 & 0.018 &  5.7E-03 & 8.6E-03 & 0.87 & 0.05 & -18.0 & 19.4 \\
A611       & 120.2345 &  36.0534 & 0.112 & 0.016 &  1.6E-02 & 8.3E-03 & 0.97 & 0.04 &  43.6 & 42.8 \\
1115+01    & 168.9666 &   1.5000 & 0.065 & 0.010 &  5.0E-03 & 5.8E-03 & 0.92 & 0.04 & -22.9 & 21.5 \\
1149+22    & 177.3988 &  22.3989 & 0.027 & 0.010 &  3.7E-03 & 4.5E-03 & 0.96 & 0.03 & -50.1 & 25.7 \\
A1423      & 179.3149 &  33.6146 & 0.099 & 0.029 &  4.1E-02 & 1.6E-02 & 0.65 & 0.14 &  61.0 & 14.9 \\
1206-08    & 181.5508 &  -8.8010 & 0.088 & 0.010 &  4.2E-03 & 3.9E-03 & 0.88 & 0.04 & -58.2 & 14.6 \\
1226+33    & 186.7416 &  33.5480 &  \nodata &  \nodata &  2.6E-02 & 8.1E-03 & 0.94 & 0.04 &  19.3 & 34.1 \\
1311-03    & 197.7554 &  -3.1775 & 0.055 & 0.012 &  2.5E-02 & 5.7E-03 & 0.94 & 0.04 &  \nodata & \nodata \\
1347-11    & 206.8783 & -11.7532 & 0.102 & 0.009 &  1.1E-02 & 3.4E-03 & 0.96 & 0.03 & -27.8 & 17.2 \\
1423+24    & 215.9509 &  24.0798 & 0.081 & 0.016 &  1.4E-02 & 7.7E-03 & 0.84 & 0.06 &  16.4 & 15.4 \\
1532+30    & 233.2231 &  30.3508 & 0.061 & 0.016 & -8.2E-03 & 8.7E-03 & 0.97 & 0.07 &  \nodata & \nodata \\
1720+35    & 260.0695 &  35.6081 & 0.054 & 0.008 &  6.3E-03 & 3.6E-03 & 0.94 & 0.03 &  89.8 &  7.7 \\
A2261      & 260.6082 &  32.1355 & 0.116 & 0.020 &  3.2E-02 & 1.0E-02 & 0.85 & 0.09 &  68.1 & 18.3 \\
1931-26    & 292.9564 & -26.5751 & 0.051 & 0.010 & -4.1E-03 & 6.5E-03 & 0.88 & 0.04 &   0.9 &  9.7 \\
2129-07    & 322.3583 &  -7.6925 & 0.081 & 0.014 &  1.1E-02 & 5.5E-03 & 0.94 & 0.03 &  29.7 & 44.2 \\
2129+00    & 322.4146 &   0.0907 & 0.097 & 0.018 & -8.0E-04 & 1.3E-02 & 0.90 & 0.06 &  28.5 & 18.5 \\
MS2137     & 325.0622 & -23.6617 & 0.105 & 0.028 & -1.8E-02 & 1.5E-02 & 0.84 & 0.12 & -80.6 & 19.6 \\
2248-44    & 342.1839 & -44.5308 & 0.091 & 0.015 &  1.4E-03 & 5.4E-03 & 0.94 & 0.04 & -62.6 & 13.7
\enddata
\tablecomments{SZE morphological parameters computed as described in Section~\ref{section:SZEmorph}. No concentration is listed for 1226+33 (the most distant in CLASH at $z\sim 0.9$) because 100 kpc is small compared to the Bolocam PSF for this cluster. Three clusters do not have well-constrained PA values, and PA therefore are not reported for those clusters.}
\end{deluxetable*}

\acknowledgments


MD was partially supported by an STScI/NASA award 
HST-GO-12065.07-A  and NASA ADAP award NNX13AI41G. AB was partially 
supported by the same NASA ADAP award NNX13AI41G and a Chandra archive grant SAO AR3-14013X.
ER was supported by FP7-PEOPLE-2013-IIF (Grant Agreement PIIF-GA-2013-627474) and NSF AST-1210973.
SE acknowledges the financial contribution from contracts ASI-INAF I/009/10/0 and PRIN-INAF 2012.
This work was supported in part by National Science Foundation Grant No. PHYS-1066293 and the hospitality of the Aspen Center for Physics.
JS was supported by NSF/AST-0838261, NASA/NNX11AB07G, and the Norris Foundation CCAT Postdoctoral Fellowship. NGC was partially supported by a NASA Graduate Student Research Fellowship. The Bolocam observations were partially supported by the Gordon and Betty Moore Foundation. This research made use of the Caltech Submillimeter Observatory, which was operated at the time by the California Institute of Technology under cooperative agreement with the National Science Foundation (NSF/AST-0838261).
GY acknowledges  financial support from   MINECO's  grant   AYA2012-31101. 
 The MUSIC simulations have been performed in the Marenostrum Supercomputer at  the Barcelona Supercomputer Center,  thanks to time granted by the Red Espa\~nola de Supercomputaci\'on.


{\it Facilities:} \facility{Chandra X-ray Observatory, Hubble Space Telescope/ACS, Hubble SpaceTelescope/WFC3, Caltech Submillimeter Observatory}

\clearpage
\vspace{1em}
\bibliography{CLASHmorph}

\clearpage

\end{document}